\documentclass[prd,twocolumn,showpacs,floatfix,amsmath,nofootinbib,amssymb,floatfix]{revtex4}
\usepackage{graphicx,color,dcolumn,booktabs,bm,multirow}
\usepackage{longtable,lscape}
\usepackage{txfonts}
 \usepackage{enumitem}
\usepackage{overpic}
\usepackage{amssymb}
\usepackage{indentfirst}
\usepackage{feynmf}   
\usepackage{slashed}  
\usepackage{cases}
\usepackage{color}
\usepackage{multirow}
\usepackage{epstopdf}
\usepackage{graphicx,color,dcolumn,booktabs,bm}

\usepackage[colorlinks,
            citecolor=blue,
            anchorcolor=red,
            menucolor=red,
            linkcolor=red,
            filecolor=red,
            runcolor=red,
            urlcolor=blue,
            frenchlinks=red]{hyperref}
\usepackage{amsmath}
\usepackage{epsfig}
\usepackage{graphicx}
\usepackage{txfonts}

\begin{document}

\title{Possible triply heavy tetraquark states in a chiral quark model}

\author{Xuejie Liu$^1$}\email[E-mail: ]{1830592517@qq.com}
\author{Yue Tan$^{2}$}\email[E-mail:]{tanyue@ycit.edu.cn}
\author{Dianyong Chen$^{1,3}$\footnote{Corresponding author}}\email[E-mail:]{chendy@seu.edu.cn}
\author{Hongxia Huang$^4$}\email[E-mail:]{hxhuang@njnu.edu.cn}
\author{Jialun Ping$^4$}\email[E-mail: ]{jlping@njnu.edu.cn}
\affiliation{$^1$School of Physics, Southeast University, Nanjing 210094, P. R. China}
\affiliation{$^2$School of Mathematics and Physics, Yancheng Institute of Technology, Yancheng, 224051,  P. R. China}
\affiliation{$^3$Lanzhou Center for Theoretical Physics, Lanzhou University, Lanzhou 730000, P. R. China}
\affiliation{$^4$Department of Physics, Nanjing Normal University, Nanjing 210023, P. R. China}

\begin{abstract}
In the present work, the triply heavy tetraquarks states $QQ\bar{Q}\bar{q}$ with $Q=(c, b)$ and $q=(u, d, s)$ with all possible quantum numbers are systematically investigated in the framework of the chiral quark model with the resonating ground method. Two kinds of structures, including the meson-meson configuration (the color-singlet channels and the hidden-color channels) and the diquark-antidiquark configuration (the color sextet-antisextet and the color triplet-antitriplet), are considered. In the considered system, several bound states are obtained for the $cc\bar{c}\bar{q^{'}}$, $bb\bar{c}\bar{q^{'}}$ and  $bc\bar{c}\bar{q}$ tetraquarks. From the present estimations, we find that the coupled channel effect is of great significance for forming the below thresholds tetraquark states, which are stable for strong decays.
\end{abstract}

\maketitle

\setcounter{totalnumber}{5}
\section{\label{sec:introduction}Introduction}

Searching for multiquark states has become one of the most important and interesting topics of hadron physics, and the experimental observations and theoretical investigations shall deepen our understanding of the nonperturbative QCD~\cite{Chen:2016qju, Swanson:2006st, Voloshin:2007dx, Chen:2016heh, Esposito:2016noz, Lebed:2016hpi, Guo:2017jvc}. At the early beginning of the quark model, the notion of multiquark states had been proposed~\cite{Gell-Mann:1964ewy}. But there had been no progress on the experimental side for a long time. A turning point came in the year of 2003, when the Belle Collaboration reported their observation of a new charmonium-like state $X(3872)$ in the exclusive  $B^{\pm}\to K^\pm \pi^+ \pi^- J/\psi $ decays~\cite{Belle:2003nnu}. Since then, a growing number of new hadron states have been observed experimentally, which attract the great interest of experimentists and theorists.

Among the new hadron states observed in the recent two decades, there are some good candidates of QCD exotic states, which can be classified into different categories according to different criteria. For example, for the charmonium-like states, we can divide them into two types according to the carried charges, i.e., the neutral and charged categories. One can also classify the new hadron states by their most possible quark components into tetraquark, pentaquark states, etc. It is interesting to notice that almost all the new hadron states have at least one heavy constituent quark or antiquark component. Since the mass of the heavy quarks is much larger than that of the light quarks, one can usually discuss the properties of hadrons with heavy quark components in the heavy quark limit. In this case, the number of heavy constituent quark/antiquark can also be used to classify the new hadron states. According to this criterion, we separate the observed new hadron states into three types, which are states with one, two, and four heavy quark/antiquark components, respectively. In the following, we select some typical examples for each type  and present a short review.

\begin{itemize}[leftmargin=*]
\item {\it States with one heavy constituent quark/antiquark.} The charmed-strange states $D_{s0}(2317)$ and $D_{s1}(2460)$ could be good examples of exotic states with one heavy constituent quark. In 2003, the BaBar Collaboration reported a narrow peak, named $D_{s0}(2317)$, in the $D_s^+ \pi^0 $ invariant mass spectrum~\cite{BaBar:2003oey}, and later this state was confirmed by  CLEO~\cite{CLEO:2003ggt} and Belle~\cite{ Belle:2003guh} Collaborations. In addition to the $D_{s0}^{\ast}(2317)$, the CLEO collaboration observed another narrow peak, named $D_{s1}(2460)$, in the $D_s^{\ast+} \pi^0$ invariant mass spectrum.~\cite{CLEO:2003ggt}. The $J^P$ quantum numbers of the $D_{s0}^\ast(2317)$ and $D_{s1}(2460)$ indicated that they could be good candidates of $1^3P_0$ and $1^3P_2$ charmed-strange mesons~\cite{Godfrey:2003kg, Rosner:2006jz, Liu:2020ruo}. However, the observed masses of $D_{s0}^\ast(2317)$ and $D_{s1}(2460)$ are much lower than the ones of the charmed strange mesons predicted by conventional quark model~\cite{Godfrey:1985xj}, which made the charmed strange mesons assignments questionable. In addition, the mass of the $D_{s0}^\ast(2317)$ and $D_{s1}(2460)$ are just several tens MeV below the thresholds of $DK$ and $D^\ast K$, thus it was natural to interpret $D_{s0}^\ast(2317)$ and $D_{s1}(2460)$ as $DK$ and $D^\ast K$ molecular states~\cite{Barnes:2003dj, Navarra:2015iea, Kolomeitsev:2003ac, Hofmann:2003je, Guo:2006fu, Zhang:2006ix, Rosner:2006vc, Guo:2006rp, Xiao:2016hoa}, respectively. Moreover, these two states can also be assigned as $c\bar{s} q\bar{q}$ tetraquark states~\cite{Cheng:2003kg,Chen:2004dy,Kim:2005gt,Nielsen:2005ia,Terasaki:2005kc,Wang:2006uba}.

\item {\it States with two heavy constituent quarks/antiquarks.} As a typical example of exotic states with two heavy constituent quarks/antiquarks, $X(3872)$ is the first charmonium-like state with the longstanding puzzle, which was observed in 2003 by the Belle Collaboration in the $\pi^{+}\pi^{-}J/\psi$ invariant mass distributions of the exclusive decay process $B^{\pm}\rightarrow K^{\pm}\pi^{+}\pi^{-}J/\psi$~\cite{Belle:2003nnu}, and then confirmed by CDF~\cite{Acosta:2003zx,Abulencia:2005zc,Abulencia:2006ma,Aaltonen:2009vj}, D0~\cite{Abazov:2004kp}, Babar~\cite{Aubert:2004fc,Aubert:2004ns,Aubert:2005eg,Aubert:2005zh,Aubert:2005vi,Aubert:2006aj,Aubert:2007rva,Aubert:2008gu,Aubert:2008ae,delAmoSanchez:2010jr} ,CMS~\cite{CMS:2011yra,Vesentini:2012lea,Chatrchyan:2013cld,DallOsso:2013rtt,DallOsso:2014cmg,Sirunyan:2020qir}, LHCb~\cite{Aaij:2011sn,LHCb:2011bia,LHCb:2011cra,Aaij:2013zoa,Aaij:2013rha,Aaij:2014ala,Aaij:2015eva,Aaij:2016kxn,Aaij:2017tzn,Aaij:2019zkm,Durham:2020zuw,Aaij:2020qga,
Aaij:2020xjx,Aaij:2020tzn} and BESIII\cite{Ablikim:2013dyn,Ablikim:2019soz,Ablikim:2019zio,Ablikim:2020xpq} in various processes. The $I(J^{PC})$ quantum numbers of $X(3872)$ have been determined to be $0(1^{++})$, which are well consistent with the ones of $\chi_{c1}(2P)$. Thus, in the literatures, $X(3872)$ was interpreted as $\chi_{c1}(2P)$ charmonium~\cite{Barnes:2003vb,Eichten:2004uh,Chen:2007vu,Meng:2007cx,Liu:2007uj,Wang:2010ej,Kalashnikova:2010hv,Wang:2012cp}. However, the measured mass is much different with the expectation of the conventional quark model~\cite{Liu:2008qb}, but sandwiched by the thresholds of $D^0 \bar{D}^{\ast 0}$ and $D^+ {D}^{\ast -}$. Moreover, the measured ratio of the branching fractions of $X(3872)\to \rho^0 J/\psi$ and $X(3872)\to \omega J/\psi$ indicated a large isospin violation~\cite{Gamermann:2009fv}, which is also inconsistent with conventional charmonium expectations. Thus the charmonium interpretations became questionable and some QCD exotic interpretations have been proposed, such as molecular~\cite{Voloshin:1976ap,DeRujula:1976zlg,Tornqvist:1993ng,Thomas:2008ja,Lee:2009hy,Chen:2009zzi,Gamermann:2009uq,Ortega:2010qq,
Guo:2013sya,Wang:2013kva,Wong:2003xk,Swanson:2003tb} and tetraquark~\cite{Vijande:2004vt,Maiani:2005pe,Ebert:2005nc,Navarra:2006nd,Cui:2006mp,Matheus:2006xi,Nielsen:2006jn,Dubnicka:2010kz,Dubnicka:2011mm, Maiani:2004vq,Wang:2013vex,Close:2003mb,
Li:2004sta,Petrov:2005tp} interpretations.

\item {\it States with four heavy constituent quarks/antiquarks.}  In the year of 2020, the first tetraquark composed of four heavy constituent quarks/antiquarks, named $X(6900)$, was observed in the di-$J/\psi$ invariant mass distributions~\cite{LHCb:2020bwg}. Subsequently, this state  was confirmed in the same channel by CMS Collaboration~\cite{CMS}, and then the existence of $X(6900)$ was verified by the ATLAS Collaboration in the di-$J/\psi$ as well as $J/\psi\psi(2S)$ invariant mass distributions~\cite{ATLAS}. Besides $X(6900)$, some additional resonance states in this energy range have been reported, such as $X(6600)$ and $X(7200)$ by CMS Collaboration~\cite{CMS}, and $X(6200)$, $X(6600)$ as well as $X(7200)$ by ATLAS Collaboration~\cite{ATLAS}. These recent experiment progress have inspired intensive theoretical investigations. The interpretations of their natures have been discussed extensively in compact tetraquark~\cite{Albuquerque:2020hio, liu:2020eha, Lu:2020cns, Giron:2020wpx, Dosch:2020hqm, Yang:2020wkh, Huang:2020dci}, $c\bar{c}$ hybrid ~\cite{Wan:2020fsk} and Higgs-like boson~\cite{Zhu:2020snb} scenarios, and the dynamical rescattering mechanism of double-charmonium channels as well~\cite{Guo:2020pvt, Dong:2020nwy, Gong:2020bmg, Wang:2020wrp, Wang:2022jmb, Gong:2022hgd, Liang:2022rew}.
\end{itemize}

The observations of the fully heavy tetraquark states makes tetraquark spectroscopy abundant and systematic. However, one can find that the tetraquark states with three heavy quark/antiquarks, i.e., $QQ\bar{Q}\bar{q}\ (q=u, d, s)$, absent experimentally. The triply heavy tetraquark states are different from the already discovered quarkonium-like states, it might in sense offer a new platform of studying the internal structure of the exotic states. On the theoretical side,  in the frame of color-magnetic interactions, the triply heavy tetraquark states were systematically investigated and some exotic tetraquark states were predicted~\cite{Chen:2016ont}. The QCD sum rule estimations indicated that the triply heavy tetraquarks states, $cc\bar{c}\bar{q}$, $cc\bar{b}\bar{q}$ and $bc\bar{b}\bar{q}$, with quantum numbers $J^{P}=0^{+}$ and $J^{P}=1^{+}$ are all heavier than the corresponding meson-meson thresholds, while the $bb\bar{b}\bar{q}$ tetraquarks were expected to be stable for strong decay~\cite{Jiang:2017tdc}. However, the estimations in the extended chromomagnetic model~\cite{Weng:2021ngd}, nonrelativistic quark model~\cite{Silvestre-Brac:1993zem}, extended relativized quark model~\cite{Lu:2021kut} indicated that there was no bound triply-heavy tetraquark state. In a word, the existence of the triply heavy tetraquark states is still an open question. In the present work, we employ a nonrelativistic chiral quark model (ChQM) to estimate the mass spectra of the $S$-wave triply heavy tetraquark states with the possible $J^P$ quantum numbers to be  $0^+$, $1^+$, $2^+$, to further check the existence of triply heavy tetraquark states.

The work is organized as follows. In Section~\ref{model} and section~\ref{model1}, the theoretical framework utilized in present estimations is presented, which includes the chiral quark model and the Resonating Group Method (RGM). Section ~\ref{results} is devoted to the analysis and discussion of the obtained results. In the last section, we give a short summary.


\section{THE CHIRAL QUARK MODEL }{\label{model}}
In the quark model, the Hamiltonian of a hadron is generally written as~\cite{Valcarce:2005em},
\begin{eqnarray}
H = \sum_{i=1}^{4} \left(m_i+\frac{\boldsymbol{p}_i^2}{2m_i}\right)-T_{CM}+\sum_{j>i=1}^4V(r_{ij}),
\end{eqnarray}
with $m_i$ and $p_i$ are the mass and momentum of the $i$th quark, respectively. $T_{CM}$ is the center-of-mass kinetic energy, which is usually subtracted without losing generality since one mainly focuses on the internal relative motions. $V(r_{ij})$ indicates the interaction potential between the $i$th and $j$th quarks.

As for the ChQM, it is constructed based on the fact that the light current quarks are nearly massless, which lead to the chiral symmetry. However, due to the interactions of the quarks with the gluon medium, the current quarks become dressed and such dressed current quarks can be approximately described by the massive constituent quarks. In practice, the masses of the constituent quarks in the ChQM are determined by reproduce the spectrum of the conventional hadrons, and this model has been widely used to investigate the study of the spectra of mesons containing heavy quarks~\cite{Segovia:2008zz,Segovia:2010zzb,Segovia:2011tb,Segovia:2016xqb}, the electromagnetic, weak and strong decays and reactions of mesons as well~\cite{Segovia:2016xqb,Segovia:2011dg,Segovia:2011zza,Segovia:2012cd,Segovia:2013kg,Segovia:2014mca}, the phenomena related to multiquark structures~\cite{Ortega:2009hj,Ortega:2016hde,Jin:2020yjn,Yan:2021glh,Liu:2020yen,Jin:2020jfc,Liu:2018nse}.
In addition, in the ChQM, the interaction potential usually includes the Goldstone-boson exchange potentials, the perturbative one-gluon interaction, and a confinement potential. Furthermore, when one only considers the $S-$wave tetraquark system, the spin-orbit and tensor contributions can be ignored, thus the two body interaction potential reads,
\begin{eqnarray}
V(r_{ij}) =V_{\mathrm{OGE}}(r_{ij})+V_{\chi}(r_{ij})+ V_{\mathrm{CON}}(r_{ij}).
\end{eqnarray}
where $V_{\mathrm{OGE}}(r_{ij})$ indicates the potential resulted from one gluon exchange, and its concrete form is
\begin{eqnarray}
V_{\mathrm{OGE}}(r_{ij})& =& \frac{1}{4}\alpha^{ij}_{s} \boldsymbol{\lambda}^{c}_i \cdot
\boldsymbol{\lambda}^{c}_j\nonumber\\
&&\times\left[\frac{1}{r_{ij}}-\frac{\pi}{2}\delta(\boldsymbol{r}_{ij})\left(\frac{1}{m^2_i}+\frac{1}{m^2	_j}
+\frac{4\boldsymbol{\sigma}_i\cdot\boldsymbol{\sigma}_j}{3m_im_j}\right)\right],
\end{eqnarray}
where $\boldsymbol{\sigma}$ and $\boldsymbol{\lambda^{c}}$ are the Pauli matrices and SU(3) color matrix, respectively. $\alpha_s^{ij}$ is the QCD-inspired scale-dependent quark-gluon coupling constant, which offers a consistent description of mesons from light to heavy-quark sectors,  and it can be determined by the mass splits between different mesons\footnote{It worth to mention that the $V_{\mathrm{OGE}}(r_{ij}) \propto 1/r_{ij}$ is very singular at short range. Similar to the case of hydrogen atom, the radial wave function should be proportional to $r_{ij}$ for the $S$-wave state. Thus, the matrix elements of $V_{\mathrm{OGE}}$ are finite.}. As for the confinement potential, the harmonic oscillator potential is adopted, which is,
\begin{equation}
 V_{\mathrm{CON}}(r_{ij}) =  -a_{c}\boldsymbol{\lambda^{c}_{i}\cdot\lambda^{c}_{j}}\left[r_{ij}^2+V_{0_{ij}}\right],
\end{equation}
where $a_{c}$ represents the strength of the confinement potential and $V_{0_{ij}}$ is the zero-point energies, which can be determined by the mass shift between different mesons.

The Goldstone-boson exchange interactions between light quarks appear because of the dynamical breaking of chiral symmetry. For the $QQ\bar{Q}\bar{q}$ with $(Q=(c, b),\ q=(u, d, s))$ systems, the $\pi$, $K$ and $\eta$ exchange interactions do not work due to the quark components. Thus in this paper, the Goldstone-boson exchange interactions are not considered.

The concrete values of these parameters are collected in Table~\ref{parameters}. In addition, the details of how to obtain these parameters can also be found in Ref.~\cite{Liu:2022vyy}. The calculated mesons masses in comparison with experimental values are shown in Table~\ref{mass}. It should be noticed that the parameters in the potentials are obtained by reproducing the mass spectra of conventional mesons, but the two-body quark-quark interaction potentials could be extended to investigate the multiquark system, where the difference between the color configurations is reflected by the product of the SU(3) color matrix $\boldsymbol{\lambda^{c}_{i}\cdot\lambda^{c}_{j}}$.

\begin{table}[htb]
\caption{The concrete values of the model parameters, which are determined by reproducing the masses of mesons listed in Table~\ref{mass}.}
\renewcommand\arraystretch{1.35}
\begin{tabular}{p{2.5cm}<\centering p{2.5cm}<\centering p{2.5cm}<\centering }
 \toprule[1pt]
      & Parameter  &Value   \\
      \midrule[1pt]
Quark masses  &$m_u$ (MeV)                      & 313 \\
              &$m_s$ (MeV)                      & 536 \\
              &$m_c$ (MeV)                      & 1728 \\
              &$m_b$ (MeV)                          & 5112 \\
      \midrule[0.5pt]
confinement   &$b$ (fm)                         &0.2\\
              &$a_{c}$ (MeV $\mathrm{fm}^{-2}$)           &101 \\
              &$V_{0_{us}}\ (\mathrm{fm}^2)$               &-3.7467\\
              &$V_{0_{uc}}\ (\mathrm{fm}^2)$               &-2.8684\\
              &$V_{0_{ub}}\ (\mathrm{fm}^2)$               &-2.6750\\
              &$V_{0_{sc}}\ (\mathrm{fm}^2)$               &-1.9211\\
              &$V_{0_{sb}}\ (\mathrm{fm}^2)$               &-1.7566\\
              &$V_{0_{cc}}\ (\mathrm{fm}^2)$               &-0.7367\\
              &$V_{0_{cb}}\ (\mathrm{fm}^2)$               &-1.0557\\
              &$V_{0_{bb}}\ (\mathrm{fm}^2)$               &2.6857\\
     \midrule[0.5pt]
OGE           &$\alpha_{s}^{us}$                     &0.0716\\
              &$\alpha_{s}^{uc}$                     &0.1127\\
              &$\alpha_{s}^{ub}$                     &0.1057\\
              &$\alpha_{s}^{sc}$                     &0.1957\\
              &$\alpha_{s}^{sb}$                     &0.1930\\
              &$\alpha_{s}^{cc}$                     &0.4953\\
              &$\alpha_{s}^{cb}$                     &0.3241\\
              &$\alpha_{s}^{bb}$                     &2.3401\\
\bottomrule[1pt]
\end{tabular}
\label{parameters}
\end{table}

\begin{table}[!htp]
\caption{The Masses (in units of MeV) of the mesons. The measured values of the masses ~\cite{ParticleDataGroup:2018ovx} are also presented for comparison.}
\renewcommand\arraystretch{1.35}
\begin{tabular}{p{1.0cm}<\centering p{0.8cm}<\centering p{0.8cm}<\centering p{0.8cm}<\centering p{0.8cm}<\centering p{0.8cm}<\centering p{0.8cm}<\centering p{0.8cm}<\centering p{0.8cm}<\centering}
\toprule[1pt]
      &$K$      &$K^{*}$ &$B$     &$B^{*}$ &$B_{s}$    &$B_{s}^{*}$  &$B_{c}$ &$B_{c}^{*}$\\ \midrule[1pt]
Expt  &495      & 892    &5280    &5325    &5366       &5415         &6275    & $...$        \\
Model  &495      & 892    &5280    &5325    &5366       &5415         &6275    &6300     \\
      &$\eta_{b}$           &$\Upsilon$   &$D$     &$D^{*}$ &$D_{s}$ &$D_{s}^{*}$ &$\eta_{c}$ &$J/\psi$ \\
Expt  &9398                 &9459         &1865    &2007    &1968    &2112        &2984       &3097     \\
Model  &9398                 &9459         &1865    &2007    &1968    &2112        &2984       &3097 \\
\bottomrule[1pt]
\end{tabular}
\label{mass}
\end{table}

\begin{figure}[htb]
\includegraphics[scale=0.45]{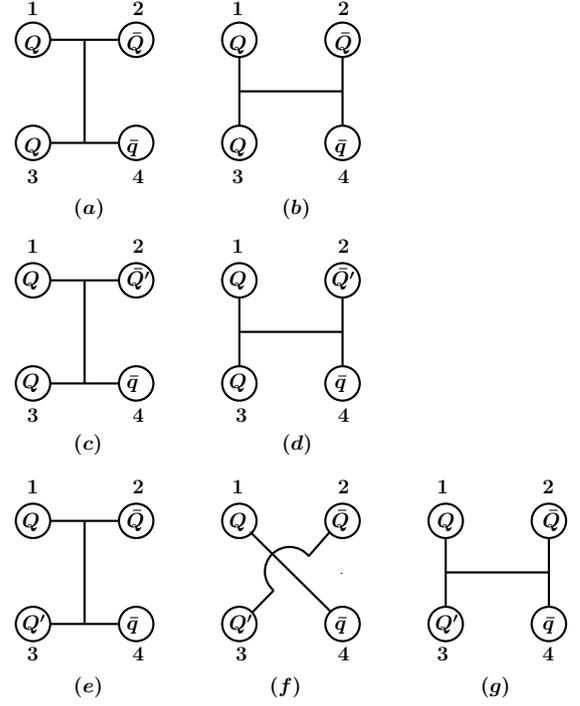}
\vspace{0.1cm} \caption{Two types of configurations in $QQ\bar{Q}\bar{q}, QQ\bar{Q^{'}}\bar{q}$ and $QQ^{'}\bar{Q}\bar{q}$ tetraquarks.
For the $QQ\bar{Q}\bar{q}$ system, there are two structures: the meson-meson configuration (diagram(a)) and the diquark-antidiquark configuration (diagram(b)). For the $QQ\bar{Q^{'}}\bar{q}$ system, diagrams (c) and (d) correspond to the meson-meson and the diquark-antidiquark configurations, respectively. For the $QQ^{'}\bar{Q}\bar{q}$ system, diagrams (e) and (f) correspond to meson-meson configuration, while diagram (g) refers to the diquark-antidiquark configuration.}
\label{fig1}
\end{figure}


\begin{table*}[htb]
\begin{center}
\caption{\label{channels} All the possible channels for different $J^P$ quantum numbers, where $[i,j,k]$ denotes the channels with $i$, $j$, and $k$ to be the indices of flavor, spin, and color, respectively.}
\begin{tabular}{p{1.5cm}<\centering  p{1.5cm}<\centering p{0.2cm}<\centering p{1.5cm}<\centering p{1.5cm}<\centering p{0.2cm}<\centering p{1.5cm}<\centering p{1.5cm}<\centering p{1.5cm}<\centering p{1.5cm}<\centering p{1.5cm}<\centering }
\toprule[1pt]
\multicolumn{2}{c}{$QQ\bar{Q}\bar{q}$} & & \multicolumn{2}{c}{$QQ\bar{Q^{\prime}}\bar{q}$}  & & \multicolumn{4}{c}{$QQ^{\prime}\bar{Q}\bar{q}$} \\
\cline{1-2} \cline{4-5} \cline{7-10}
$J^{P}$  & Channel  & & $J^{P}$  & Channel &&$J^{P}$  & Channel  & $J^{P}$  & Channel\\
\midrule[1pt]
$0^{+}$                & [1,1,1] & &                $0^{+}$  &[3,1,1] &&                $0^{+}$  &[5,1,1] &$1^{+}$&[6,3,1] \\
                       & [1,1,2] & &                         &[3,1,2] &&                         &[5,1,2] &      &[6,3,2]  \\
                       & [1,2,1] & &                         &[3,2,1] &&                         &[5,2,1] &      &[6,4,1] \\                                   & [1,2,2] & &                         &[3,2,2] &&                         &[5,2,2] &      &[6,4,2]\\
                       & [2,1,3] & &                         &[4,1,3] &&                         &[6,1,1] &      &[6,5,1] \\                                   & [2,2,4] & &                         &[4,2,4] &&                         &[6,1,2] &      &[6,5,2] \\
$1^{+}$                & [1,3,1] & &                $1^{+}$  &[3,3,1] &&                         &[6,2,1] &      &[7,3,3]\\
                        & [1,3,2]& &                         &[3,3,2] &&                         &[6,2,2] &      &[7,3,4]\\
                        & [1,4,1]& &                         &[3,4,1] &&                         &[7,1,3] &      &[7,4,3]\\
                        & [1,4,2]& &                         &[3,4,2] &&                         &[7,1,4] &      &[7,4,4] \\
                        & [1,5,1]& &                         &[3,5,1] &&                         &[7,2,3] &      &[7,5,3]\\
                        & [1,5,2]& &                         &[3,5,2] &&                         &[7,2,4] &      &[7,5,4]\\
                        & [2,3,3]& &                         &[4,3,3] &&                $1^{+}$  &[5,3,1] & $2^{+}$&[5,6,1]\\
                        & [2,4,4]& &                         &[4,4,4] &&                         &[5,3,2] & &[5,6,2]\\
                        & [2,5,4]& &                         &[4,5,4] &&                         &[5,4,1] & &[6,6,1]\\
$2^{+}$                 & [1,6,1]& &                $2^{+}$  &[3,6,1] &&                         &[5,4,2] & &[6,6,2]\\
                        & [1,6,2]& &                         &[3,6,2] &&                         &[5,5,1] & &[7,6,3]\\
                        & [2,6,4]& &                         &[4,6,4] &&                         &[5,5,2] & &[7,6,4]\\
\bottomrule[1pt]
\end{tabular}
\end{center}
\label{channe}
\end{table*}


\section{THE RESONATING GROUP METHOD }{\label{model1}}
In the present work, the triply heavy tetraquark systems are estimated by using the resonating group method~\cite{Kamimura:1981oxj}. In this method, the multiquark system can be divided into two clusters, which are frozen inside, so one only needs to consider the relative motion between the two clusters. The conventional ansatz for two-cluster (cluster A and B) wave functions is,
\begin{eqnarray}\label{wave2}
  \psi_{4q} &=&  \mathcal{A}\left[\left[\psi_{A}(\boldsymbol{\rho}_A)\psi_{B}(\boldsymbol{\rho}_B)\right]^{[\sigma]IS}\otimes\chi_{L}(\textbf{R})\right]^{J},
\end{eqnarray}
where $\mathcal{A}$ is the antisymmetry operator of triply heavy tetraquarks.

For $QQ\bar{Q}\bar{q}$ system, one has,
\begin{equation}
    \mathcal{A}=1-P_{13}.
 \end{equation}
this antisymmetry operator becomes,
 \begin{equation}
    \mathcal{A}=1-P_{13},
 \end{equation}
for $QQ\bar{Q^{'}}\bar{q}$ system, and for $QQ^{'}\bar{Q}\bar{q}$ system, due to the absence of any homogeneous quarks, then antisymmetry operator becomes a unit operator, which is,
 \begin{equation}
    \mathcal{A}=1.
 \end{equation}
 Moreover, $[\sigma]=[222]$ gives the total color symmetry, and $I, S, L$ and $J$ represent flavor, spin, orbital and total angular momenta, respectively. $\psi_{A}$ and $\psi_{B}$ are the two-quark cluster wave functions, which are,
\begin{eqnarray}
 \psi_{A} &=& \left(\frac{1}{2\pi b^{2}}\right)^{3/4}  e^{-\boldsymbol{\rho_{A}}^{2}/(4b^2)} \eta_{I_{A}}S_{A}\chi_{A}^{c} ,\\
  \psi_{B} &=& \left(\frac{1}{2\pi b^{2}}\right)^{3/4} e^{-\boldsymbol{\rho_{B}}^{2}/(4b^2)} \eta_{I_{B}}S_{B}\chi_{B}^{c} ,
\end{eqnarray}
where $\eta_I$, $S$, and $\chi$ represent the flavor, spin, and internal color terms of the cluster wave functions, respectively. According to Fig.~\ref{fig1}, we define different Jacobi coordinates for different diagrams. As for the meson-meson configuration in Fig.~\ref{fig1}, the Jacobi coordinates are,
\begin{eqnarray}\label{wave6}
\boldsymbol{\rho_{A}}&=&\boldsymbol{r_{q_1}-r_{\bar{q}_2}},  \ \ \ \ \boldsymbol{\rho_{B}}=\boldsymbol{r_{q_3}-r_{\bar{q}_4}},\nonumber \\
\boldsymbol{R_{A}}&=& \frac{m_{1}\boldsymbol{r_{q_1}}+m_{2}\boldsymbol{r_{\bar{q}_2}}}{m_{1}+m_{2}},\nonumber \\
 \boldsymbol{R_{B}}&=& \frac{m_{3}\boldsymbol{r_{q_3}}+m_{4}\boldsymbol{r_{\bar{q}_4}}}{m_{3}+m_{4}},\nonumber \\
\boldsymbol{R}&=&\boldsymbol{R_{A}-R_{B}}, \nonumber \\ \boldsymbol{R_{c}}&=&\frac{m_{1}\boldsymbol{r_{q_1}}+m_{2}\boldsymbol{r_{\bar{q}_2}}+m_{3}\boldsymbol{r_{q_3}}+m_{4}\boldsymbol{r_{\bar{q}_4}}}{m_{1}+m_{2}+m_{3}+m_{4}}.
\end{eqnarray}
where the subscript $q/\bar{q}$ indicates the quark/antiquark particle, while the number indicates the quark position in Fig.~\ref{fig1}. By interchanging $\boldsymbol{r}_{q_1}$ with $\boldsymbol{r}_{q_3}$, one can obtain the Jacobi coordinates in Fig.~\ref{fig1}-(f). As for the diquark-antidiquark configuration, one can also obtain the Jacobi coordinates corresponding to the diagrams in Fig.~\ref{fig1} by interchanging $\boldsymbol{r}_{q_3}$ with $\boldsymbol{r}_{\bar{q}_2}$.

From the variational principle, after variation with respect to the relative motion wave function $\chi\boldsymbol(R)=\sum_{L}\chi_{L}\boldsymbol(R)$, one obtains the RGM equation, which is,
\begin{eqnarray}\label{wave3}
  \int H\left(\boldsymbol{R, R^{\prime}}\right)\chi\left(\boldsymbol{R^{\prime}}\right)d\boldsymbol\left(R^{\prime}\right)=E \int N\left(\boldsymbol{R, R^{\prime}}\right) \chi\left(\boldsymbol{R^{\prime}}\right)d\boldsymbol\left(R^{\prime}\right),\nonumber \\
\end{eqnarray}
with $H(\boldsymbol{R, R^{\prime}})$ and $N(\boldsymbol{R, R^{\prime}})$ to be the Hamiltonian and normalization kernels, respectively. The eigenenergy $E$ and the wave functions are obtained by solving the above RGM equation. In the present estimation, the function $\chi (\boldsymbol{R})$ can be expanded by gaussian bases,  which is,
\begin{eqnarray}\label{wave5}
\chi\boldsymbol{(R)}&=&\frac{1}{\sqrt{4\pi}}\sum_{L}\left(\frac{1}{\pi b^2}\right)^{3/4}\sum_{i}^{n}C_{i,L} \nonumber\\
&&\times\int e^{-\frac{1}{2}\boldsymbol(R-S_{i})^{2}/b^2} Y^L\left(\hat{\boldsymbol{S_{i}}}\right)d\hat{\boldsymbol{S_{i}}}
\end{eqnarray}
where $C_{i,L}$ is the expansion coefficient, and $n$ is the number of gaussian bases, which is determined by the stability of the results. $\boldsymbol{S_{i}}$ is the separation of two reference centers. $\boldsymbol{R}$ is the dynamic coordinate defined in Eq.~(\ref{wave6}). After including the motion of the center of mass, i.e., 
\begin{equation}
  \phi_{C}(\boldsymbol{R_{c}})=\left(\frac{4}{\pi b^2}\right)^{3/4}\mathrm{e}^{\frac{-2 \boldsymbol{R_{c}}^{2}}{b^{2}}}.
\end{equation}
With the above formula, one can rewrite the wave function in Eq.~(\ref{wave2}) as,
\begin{eqnarray}\label{wave4}
\psi_{4q}&=&\mathcal{A} \sum_{i,L}C_{i, L}\int \frac{d \hat{\boldsymbol{S_{i}}}}{\sqrt{4 \pi}} \prod_{\alpha=1}^{2}\phi_{\alpha}\left(\boldsymbol{S_{i}}\right)\prod_{\alpha=3}^{4}\phi_{\beta}\left(\boldsymbol{-S_{i}}\right) \nonumber \\
&&\times \left[\left[\eta_{I_{A}S_{A}}\eta_{I_{B}S_{B}}\right]^{IS}Y^{L}(\hat{\boldsymbol{S_{i}}})\right]^{J}\left[\chi_{A}^{c}\chi_{B}^{c}\right]^{[\sigma]} ,
\end{eqnarray}
where $\phi_{\alpha}(\boldsymbol{S_{i}})$ and $\phi_{\beta}(\boldsymbol{-S_{i}})$ are the single-particle orbital wave functions with different reference centers, which are,
\begin{eqnarray}\label{wave1}
\phi_\alpha(\boldsymbol {S_{i}})=\left(\frac{1}{\pi
b^2}\right)^{\frac{3}{4}}e^ {-\frac{(\boldsymbol {r_{\alpha}}-\frac{1}{2}\boldsymbol
{S_i})^2}{2b^2}},
 \nonumber\\
\phi_\beta(-\boldsymbol {S_{i}})=\left(\frac{1}{\pi
b^2}\right)^{\frac{3}{4}}e^ {-\frac{(\boldsymbol {r_{\beta}}+\frac{1}{2}\boldsymbol
{S_i})^2}{2b^2}}.
\end{eqnarray}

With the reformulated ansatz as shown in Eq.~(\ref{wave4}), the RGM equation becomes an algebraic eigenvalue equation, which is,
\begin{eqnarray}
  \sum_{j,L}C_{J,L}H_{i,j}^{L,L^{\prime}} &=& E\sum_{j}C_{j,L^{\prime}}N_{i,j}^{L^{\prime}},
\end{eqnarray}
with $N_{i,j}^{L^{\prime}}$ and $H_{i,j}^{L,L^{\prime}}$ to be the  overlap of the wave functions and the matrix elements of the Hamiltonian, respectively. By solving the generalized eigenvalue problem, we can obtain the energies of the tetraquark systems $E$ and the corresponding expansion coefficient $C_{j,L}$. Finally, the relative motion wave function between two clusters can be obtained by substituting the $C_{j,L}$ into Eq.~(\ref{wave5}).

Besides the space part, we present the flavor, spin, and color parts of the wave function in Appendix-\ref{Sec:Appendix}. It is worth noting that after applying the antisymmetry operator, some wave functions may vanish, which means that some states are forbidden. For example, for the $cc\bar{b}\bar{q}$ system with $J^{P}=0^{+}$, when considering the diquark-antidiquark structure with the spin wave function forced to choose $S^{1}_{0}$, the color wave function $\chi^{c}_{3}$ would be excluded due to the constraints that the total wave function must be antisymmetric.

\begin{table*}[t]
\begin{center}
\caption{\label{cccq} The lowest-lying eigenenergies of the $cc\bar{c}\bar{n} \ n=\{u,d\}$ and $cc\bar{c}\bar{s}$ tetraquarks in the ChQM.}
\renewcommand\arraystretch{1.35}
\begin{tabular}{p{1.25cm}<\centering p{1.35cm}<\centering p{1.5cm}<\centering p{1.35cm}<\centering p{1.35cm}<\centering p{1.35cm}<\centering p{0.5cm}<\centering p{1.5cm}<\centering p{1.35cm}<\centering p{1.5cm}<\centering p{1.35cm}<\centering p{1.35cm}<\centering p{1.35cm}<\centering}
\toprule[1pt]
\multirow{2}*{$J^P$}&\multirow{2}*{$[i,j,k]$}    &\multicolumn{4}{c}{$cc\bar{c}\bar{n}$}  & & \multicolumn{4}{c}{$cc\bar{c}\bar{s}$}   \\
\cline{3-6} \cline{8-11}
& ~~&Channel &$E_{th}$ & $E_{sc}$  & $P(\%)$        &  &Channel &$E_{th}$ &$E_{sc}$  &$P(\%)$ \\
\midrule[1pt]
\multirow{9}*{$0^{+}$} &$[1,1,1]$ &$(\eta_{c}D^{})^{1}$              &4849  &4851 &$99.91\%$ && $(\eta_{c}D_{s}^{+})^{1}$     &4952 &4954&$99.98\%$\\
&$[1,2,1]$ &$(J/\psi D^{{\ast}})^{1}$          &5104  &5106 &$0.01\%$  && $(J/\psi D_{s}^{\ast+})^{1}$  &5209 &5210&$\sim0\%$ \\
&$[1,1,2]$ &$(\eta_{c}D^{})^{8}$               &      &5550 &$0.01\%$  && $(\eta_{c}D_{s}^{+})^{8}$     &     &5640&$\sim0\%$\\
&$[1,2,2]$ &$(J/\psi D^{\ast})^{8}$            &      &5563 &$0.03\%$  && $(J/\psi D_{s}^{\ast+})^{8}$  &     &5614&$\sim0\%$\\
&$[2,1,3]$ &$(cc)(\bar{c}\bar{n})$             &      &5624 &$0.01\%$  && $(cc)(\bar{c}\bar{s})$        &     &5697&$\sim0\%$ \\
&$[2,2,4]$ &$(cc)(\bar{c}\bar{n})$             &      &5421 &$0.03\%$  && $(cc)(\bar{c}\bar{s})$        &     &5498&$\sim0\%$  \\
&$E_{CC1}$ &                                   &      &4851 &          &&                               &     &4954&     \\
&$E_{CC2}$ &                                   &      &5415 &          &&                               &     &5487&                                    \\
&$E_{CC}$ &                                    &      &4851 &          &&                               &     &4954&                                \\
\midrule[1pt]
\multirow{12}*{$1^{+}$}&$[1,3,1]$ &$(\eta_{c}D^{*})^{1}$         &4991 &4993&$1.68\%$  &&$(\eta_{c}D_{s}^{*+})^{1}$      &5096 &5098&$\sim0\%$  \\
&$[1,4,1]$ &$(J/\psi   D^{})^{1}$         &4962 &4964&$96.24\%$ &&$(J/\psi   D_{s}^{+})^{1}$      &5065 &5067&$99.91\%$  \\
&$[1,5,1]$ &$(J/\psi  D^{\ast})^{1}$      &5104 &5106&$0.19\%$  &&$(J/\psi  D_{s}^{\ast+})^{1}$   &5209 &5211&$\sim0\%$  \\
&$[1,3,2]$ &$(\eta_{c}D^{*})^{8}$         &     &5522&$0.05\%$  &&$(\eta_{c}D_{s}^{*+})^{8}$      &     &5610&$\sim0\%$   \\
&$[1,4,2]$ &$(J/\psi   D^{})^{8}$         &     &5526&$0.09\%$  &&$(J/\psi   D_{s}^{+})^{8}$      &     &5614&$\sim0\%$    \\
&$[1,5,2]$ &$(J/\psi  D^{\ast})^{8}$      &     &5518&$0.65\%$  &&$(J/\psi  D_{s}^{\ast+})^{8}$   &     &5585&$\sim0\%$\\
&$[2,3,3]$ &$(cc)(\bar{c}\bar{n})$&       &5588&$0.37\%$  &&$(cc)(\bar{c}\bar{s})$          &     &5661&$\sim0\%$   \\
&$[2,4,4]$ &$(cc)(\bar{c}\bar{n})$&       &5364&$0.06\%$  &&$(cc)(\bar{c}\bar{s})$          &     &5445&$\sim0\%$    \\
&$[2,5,3]$ &$(cc)(\bar{c}\bar{n})$&       &5428&$0.17\%$  &&$(cc)(\bar{c}\bar{s})$          &     &5508&$\sim0\%$   \\
&$E_{CC1}$ &                               &     &4964&          &&                                &     &5067  &\\
&$E_{CC2}$ &                               &     &5363&          &&                                &     &5442  & \\
&$E_{CC}$ &                                &     &4963&          &&                                &     &5066  &\\
\midrule[1pt]
\multirow{6}*{$2^{+}$}&$[1,6,1]$ &$(J/\psi  D^{\ast})^{1}$       &5104 &5106&$75.23\%$ &&$(J/\psi  D_{s}^{\ast+})^{1}$  &5209 &5211&$99.79\%$  \\
&$[1,6,2]$ &$(J/\psi  D^{\ast})^{8}$       &     &5494&$11.70\%$ &&$(J/\psi  D_{s}^{\ast+})^{8}$  &     &5598&$\sim0\%$   \\
&$[2,6,4]$ &$(cc)(\bar{c}\bar{n})$ &     &5442&$13.06\%$ &&$(cc)(\bar{c}\bar{s})$         &     &5526&$\sim0\%$\\
&$E_{CC1}$ &                                &     &5106&          &&                               &     &5211  &    \\
&$E_{CC2}$ &                                &     &5442&          &&                               &     &5526  &     \\
&$E_{CC}$  &                                &     &5095&          &&                               &     &5211  &      \\
\bottomrule[1pt]
\end{tabular}
\end{center}
\end{table*}

\begin{table*}[!htb]
\begin{center}
\caption{\label{bbbq} The lowest-lying eigenenergies of the $bb\bar{b}\bar{n}$ $n=\{u, d\}$ and $bb\bar{b}\bar{s}$ tetraquarks in the ChQM.}
\renewcommand\arraystretch{1.35}
\begin{tabular}{p{1.25cm}<\centering p{1.35cm}<\centering p{1.5cm}<\centering p{1.35cm}<\centering p{1.35cm}<\centering p{1.35cm}<\centering p{0.5cm}<\centering p{1.5cm}<\centering p{1.35cm}<\centering p{1.5cm}<\centering p{1.35cm}<\centering p{1.35cm}<\centering p{1.35cm}<\centering}
\toprule[1pt]
\multirow{2}*{$J^{P}$} &\multirow{2}*{[i,j,k]}    &\multicolumn{4}{c}{$bb\bar{b}\bar{n}$}  & & \multicolumn{4}{c}{$bb\bar{b}\bar{s}$}   \\
\cline{3-6} \cline{8-11}
 & ~~&Channel &$E_{th}$ & $E_{sc}$  & $P(\%)$        &  &Channel &$E_{th}$ &$E_{sc}$  &$P(\%)$ \\
\midrule[1pt]
\multirow{9}*{$0^{+}$}&$[1,1,1]$ &$(\eta_{b}\bar{B})^{1}$                      &14679  &14681  &$99.99\%$ &&$(\eta_{b}\bar{B_{s}})^{1}$          &14766&14767 &$99.99\%$\\
                                 &$[1,2,1]$ &$(\Upsilon\bar{B}^{\ast})^{1}$    &14785  &14787  &$0.01\%$  &&$(\Upsilon \bar{B_{s}^{\ast}})^{1}$  &14875&14876 &$\sim0\%$ \\
                                 &$[1,1,2]$ &$(\eta_{b}\bar{B})^{8}$           &       &15302  &$\sim0\%$ &&$(\eta_{b}\bar{B_{s}})^{8}$          &     &15315 &$\sim0\%$  \\
                                 &$[1,2,2]$ &$(\Upsilon\bar{B}^{\ast})^{8}$    &       &15342  &$\sim0\%$ &&$(\Upsilon \bar{B_{s}^{\ast}})^{8}$  &     &15327 &$\sim0\%$ \\
                                 &$[2,1,3]$ &$(bb)(\bar{b}\bar{n})$            &       &15359  &$\sim0\%$ &&$(bb)(\bar{b}\bar{s})$               &     &15358 &$\sim0\%$  \\
                                 &$[2,2,4]$ &$(bb)(\bar{b}\bar{n})$            &       &15144  &$\sim0\%$ &&$(bb)(\bar{b}\bar{s})$               &     &15171 &$\sim0\%$  \\
                                 &$E_{CC1}$ &                                  &       &14681  &          &&                                     &     &14767 &   \\
                                 &$E_{CC2}$ &                                  &       &15143  &          &&                                     &     &15170 &   \\
                                 &$E_{CC}$ &                                   &       &14680  &          &&                                     &     &14767 &  \\
\midrule[1pt]
\multirow{12}*{$1^{+}$}               &$[1,3,1]$ &$(\eta_{b}\bar{B^{\ast}})^{1}$           &14724&14726  &$99.98\%$ &&$(\eta_{b}\bar{B_{s}^{\ast}})^{1}$    &14814&14815  &99.97\%  \\
                                             &$[1,4,1]$ &$(\Upsilon    \bar{B})^{1}$       &14740&14742  &$0.01\%$  &&$(\Upsilon    \bar{B_{s}})^{1}$       &14827&14828  &$\sim0\%$  \\
                                             &$[1,5,1]$ &$(\Upsilon    \bar{B}^{\ast})^{1}$&14785&14787  &$0.01\%$  &&$(\Upsilon    \bar{B_{s}^{\ast}})^{1}$&14875&14876  &$\sim0\%$  \\
                                             &$[1,3,2]$ &$(\eta_{b}\bar{B^{\ast}})^{8}$    &     &15292  &$\sim0\%$ &&$(\eta_{b}\bar{B_{s}^{\ast}})^{8}$    &     &15304  &$\sim0\%$ \\
                                             &$[1,4,2]$ &$(\Upsilon    \bar{B})^{8}$       &     &15294  &$\sim0\%$ &&$(\Upsilon    \bar{B_{s}})^{8}$       &     &15307  &$\sim0\%$   \\
                                             &$[1,5,2]$ &$(\Upsilon    \bar{B}^{\ast})^{8}$&     &15306  &$\sim0\%$ &&$(\Upsilon    \bar{B_{s}^{\ast}})^{8}$&     &15304  &$\sim0\%$  \\
                                             &$[2,3,3]$ &$(bb)(\bar{b}\bar{n})$   &     &15348  &$\sim0\%$ &&$(bb)(\bar{b}\bar{s})$                &     &15346  &$\sim0\%$  \\
                                             &$[2,4,4]$ &$(bb)(\bar{b}\bar{n})$   &     &15127  &$\sim0\%$ &&$(bb)(\bar{b}\bar{s})$                &     &15155  &$\sim0\%$   \\
                                             &$[2,5,3]$ &$(bb)(\bar{b}\bar{n})$   &     &15147  &$\sim0\%$ &&$(bb)(\bar{b}\bar{s})$                &     &15175  &$\sim0\%$   \\
                                             &$E_{CC1}$ &                                  &     &14726  &          &&                                      &     &14815  &\\
                                             &$E_{CC2}$ &                                  &     &15127  &          &&                                      &     &15155  & \\
                                             &$E_{CC}$ &                                   &     &14726  &          &&                                      &     &14815  &\\
\midrule[1pt]
\multirow{6}*{$2^{+}$}          &$[1,6,1]$ &$(\Upsilon    \bar{B}^{\ast})^{1}$&14785&14787  &99.99\%   &&$(\Upsilon    \bar{B_{s}^{\ast}})^{1}$ &14875&14876  &99.99\% \\
                                &$[1,6,2]$ &$(\Upsilon    \bar{B}^{\ast})^{8}$&     &15273  &$\sim0\%$ &&$(\Upsilon    \bar{B_{s}^{\ast}})^{8}$ &     &15298  &$\sim0\%$  \\
                                &$[2,6,4]$ &$(bb)(\bar{b}\bar{n})$            &     &15153  &0.01\%    &&$(bb)(\bar{b}\bar{s})$                 &     &15183  &$\sim0\%$ \\
                                &$E_{CC1}$ &                                  &     &14787&            &&                                       &     &14876  &    \\
                                &$E_{CC2}$ &                                  &     &15153&            &&                                       &     &15183  &     \\
                                &$E_{CC}$ &                                   &     &14787&            &&                                       &     &14876  &      \\
\bottomrule[1pt]
\end{tabular}
\end{center}
\end{table*}

\begin{table*}[ht]
\begin{center}
\caption{\label{part1} The average values of each operator in the Hamiltonian of the $cc\bar{c}\bar{n}$ and $bb\bar{b}\bar{n}$ tetraquark system in unit of MeV. $E_{M(J/\psi D^\ast)}$ and $E_{M(\Upsilon  B^\ast)}$ stand for the sum of the theoretical thresholds of $J/\psi D^\ast$ and $\Upsilon B^\ast$ channel, where the distance between two mesons are very large and the interactions between them are ignored. }
 \renewcommand\arraystretch{1.5}
  \begin{tabular}{p{2.cm}<\centering p{1.5cm}<\centering p{1.5cm}<\centering p{1.5cm}<\centering p{1.5cm}<\centering <\centering p{0.5cm}<\centering p{1.2cm}<\centering p{1.5cm}<\centering p{1.5cm}<\centering p{1.5cm}<\centering}
 \toprule[1pt]
     & & $\langle H_{T}\rangle$ & $\langle V_{CON}\rangle$ & $\langle V_{OGE}\rangle $  & & & $\langle H_{T}\rangle $ & $\langle V_{CON}\rangle $ & $\langle V_{OGE} \rangle$ \\
    \cline{2-5} \cline{7-10}
  \multirow{9}*{$J^{P}=2^{+}$}  & $E_{(J/\psi  D^{\ast})^{1}}$   &1802.8 &-1812.7 &-380.4 &&$E_{(\Upsilon \bar{B}^{\ast})^{1}}$  &1382.6 &-1328.1 &-952.4 \\
   &$E_{CC1}$                       &1802.5 &-1812.6 &-380.3 && $E_{CC1}$                            &1382.5 &-1328.1 &-952.4 \\
   &$E_{CC2}$                       &1994.3 &-1695.1 &-353.3 & &$E_{CC2}$                            &1594.4 &-1255.3 &-834.8  \\
   &$E_{CC}$                        &1801.2 &-1812.7 &-390.5 && $E_{CC}$                             &1382.3 &-1328.0 &-952.3  \\
   &$E_{M(J/\psi  D^{\ast})}$       &1800.1 &-1812.8 &-380.5 &&$E_{M(\Upsilon \bar{B}^{\ast} )}$             &1380.5 &-1328.9 &-953.5 \\
   &$\Delta E_{(J/\psi  D^{\ast})^{1}}$   &2.7    &0.1     &0.1  &&$\Delta E_{(\Upsilon \bar{B}^{\ast})^{1}}$ &2.6  &0.8     &1.1\\
   &$\Delta  E_{CC1}$                     &2.4    &0.2     &0.2  &&$\Delta  E_{CC1}$                         &2.0  &0.8     &1.1\\
   &$\Delta  E_{CC2}$                     &194.2  &117.7   &27.2 &&$\Delta  E_{CC2}$                         &213.9&73.6    &118.7\\
   &$\Delta  E_{CC}$                      &1.1    &0.1     &-10.0&&$\Delta  E_{CC}$                          &1.8  &0.9     &1.3\\
  \bottomrule[1pt]
\end{tabular}
\end{center}
\end{table*}

\section{RESULTS AND DISCUSSIONS}{\label{results}}
In the present calculation, the triply heavy tetraquark systems are evaluated by taking into account the meson-meson and diquark-antidiquark configurations in the ChQM, which have been shown in Fig~\ref{fig1}. To exhaust all possible configurations of the $QQ\bar{Q}\bar{q}$ systems, we divide them into three classes, which are, the $QQ\bar{Q}\bar{q}$ system including $cc\bar{c}\bar{q}$ and $bb\bar{b}\bar{q}$, the $QQ\bar{Q^{\prime}}\bar{q}$ system including $cc\bar{b}\bar{q}$ and $bb\bar{c}\bar{q}$, and the $QQ^{\prime}\bar{Q}\bar{q}$ system including $cb\bar{b}\bar{q}$ and $cb\bar{c}\bar{q}$. Moreover, in the present work, only the $S-$wave triply-heavy tetraquark states are evaluated, which indicates that the total orbital angular momenta $L$ is equal to zero. Then, the total angular momentum, $J$, coincides with the total spin, $S$, and can take values of 0, 1, and 2, then the possible $J^P$ quantum numbers of the tetraquark states could be $0^+$, $1^+$, and $2^+$. All the possible channels would be considered through the symmetry of the wave functions and all the allowed channels are listed in Table~\ref{channels}. From Table~\ref{channels}, one can find that in the ChQM the color singlet-singlet $(1_{c}\times1_{c})$ and the color octet-octet $(8_{c}\times8_{c})$ structure have been taken into account for the meson-meson configuration. Moreover, for the diquark-antidiquark configuration, both antitriplet-triplet $(\bar{3}_{c}\times3_{c})$ and sextet-antisextet $(6_{c}\times\bar{6}_{c})$  color structures have also been considered.

Our estimations of the eigenenergies of the triply tetraquark states are presented in Tables~\ref{cccq}-\ref{bcbq}. In these tables, all the allowed meson-meson and diquark-antidiquark configurations are listed. In the meson-meson channels, $(M_1 M_2)^1$ and $(M_1M_2)^8$ indicate the color singlet-singlet $(1_c \times 1_c)$ and the color octet-octet $(8_c \times 8_c)$ structures, respectively. $E_{th}$ is the experimental value of the thresholds for the physical channels. In the present work, the single-channel and channel-coupling calculations are all considered, and $E_{sc}, E_{CC1}, E_{CC2}$ and $E_{CC}$ are the estimated values of the eigenenergies of every single channel, the coupled channel for the meson-meson configurations, the coupled channel for the diquark-antidiquark configurations, and the one estimated by simultaneously considering the meson-meson and diquark-antidiquark configurations, respectively. $P$ indicates the percentages of each channel for the lowest-lying eigenenergies $E_{CC}$.

\subsection{The $QQ\bar{Q}\bar{q}$ systems}

Our estimations for the $cc\bar{c}\bar{q}$ tetraquark system are presented in Table \ref{cccq}. For the case of  $J^{P}=0^{+}$, one can find there are four channels in the meson-meson configurations, and two channels in the diquark-antidiquark configurations. For the $cc\bar{c}\bar{n}, \ n=\{u,d\}$ tetraquark states, the lowest threshold of the physical channel is 4849 MeV, which is the threshold of $\eta_c D$. Form the table, one can find the eigenenergies of every single channel in both the meson-meson and diquark-antidiquark configurations are all above the lowest threshold of the allowed physics channel, which indicates that all these tetraquark states can decay into $\eta_c D$. When one couple all the channels in a certain configuration, one can find the estimated eigenenergies are 4851 and 5415 MeV for the meson-meson and diquark-antidiquark configurations, respectively, which is still a bit higher than the threshold of $\eta_c D$. After considering both the meson-meson and diquark-antidiquark configurations simultaneously, we find the eigenenergy of $cc\bar{c} \bar{n}$ tetraquark state is about 4851 MeV, which is about 2 MeV above the threshold of $\eta_c D$. As for $cc\bar{c}\bar{s}$ tetraquark states with $J^P=0^+$, the lowest threshold of the physical channel is $4952$ MeV, which is the threshold of $\eta_c D_s^+$. As for $cc\bar{c} \bar{s}$ tetraquark state with $J^P=0^+$, the  single channel estimations show that all the tetraquark states are heavier than $\eta_c D_s^+$. The eigenenergies of the coupled-channel estimations in meson-meson and diquark-antidiquark configurations are 4954 MeV and 5487 MeV, respectively, which are all above the threshold of $\eta_c D_s^+$. Moreover, the full coupled-channel estimations, i.e., considering the meson-meson and diquark-antidiquark configurations simultaneously, indicate the eigenenergy of the $cc\bar{c}\bar{s}$ tetraquark state is 4594 MeV, which indicates that in this case, the effects of channel coupling is rather weak. It is worth noting that in the single-channel estimation the eigenenergy for the lowest physical meson-meson channel is several hundred MeV below the ones of other channels, thus, in the coupled-channel estimations, the mixings between different channels are expected to be small due to the large eigenenergy splittings.

As for the $cc\bar{c}\bar{q}$ tetraquark system with $J^P=1^+$, there are nine channels in this case, which include three color singlet channels and three hidden color channels in the meson-meson configuration, while there are three channels in the diquark-antidiquark configuration. The lowest physical meson-meson threshold is the one of $J/\psi D$, which is 4962 MeV. In the single channel estimations, no bound state is found. The eigenenergies estimated in the coupled-channel estimations of the meson-meson and diquark-antidiquark configurations are 4964 and 5363 MeV, respectively, which are all above the threshold of $J/\psi D$. By considering both the meson-meson and diquark-antidiquark configurations simultaneously, the eigenenergy of the tetraquark state with $J^P=1^{+}$ is estimated to be 4963 MeV, and the effect of the channel coupling is rather weak, which is similar to the case of $J^P=0^+$. As for the $cc\bar{c} \bar{s}$ tetraquark system, the lowest physical threshold is the one of $J/\psi D_s^+$, which is 5065 MeV. Similar to the case of $cc\bar{c}\bar{n}$ system, the eigenenergies obtained in the single channel are all above the threshold of $J/\psi D_s^+$. In addition, when we consider the channel coupling in the meson-meson and diquark-antidiquark configurations individually, the eigenenergies of the tetraquark state are estimated to be 5067 and 5442 MeV. After considering the meson-meson and diquark-antidiquark configurations simultaneously, we obtain the eigenenergy of $cc\bar{c}\bar{s}$ tetraquark state with $J^P=1^+$ is 5066 MeV, which is still a bit higher than the threshold of $J/\psi D_s^+$.

For the case of $ccc\bar{n}$ tetraquark states with $J^P=2^+$, there are two channels in the meson-meson configuration and only one channel in the diquark-antidiquark channel. The physical meson-meson threshold is 5104 MeV. Our single channel estimations indicate that the eigenenergies are all above the threshold of $J/\psi D^\ast$, and after considering the channel coupling in the meson-meson configuration, the eigenenergy is estimated to be 5106 MeV, which is still above the threshold of $J/\psi D^\ast$. When we include the meson-meson and diquark-antidiquark configuration simultaneously, the eigenenergy is estimated to be 5095 MeV, which is about 9 MeV below the threshold of $J/\psi D^\ast$ and then this tetraquark state can not decay into $J/\psi D^\ast$. Moreover, our estimations indicate in this states the dominant component is $J/\psi D^\ast$, which is about $75\%$, while the fractions of the hidden color channel, $(J/\psi D^{{\ast}})^{8}$,  and the diquark-antidiquark channel, $(cc)(\bar{c}\bar{n})$, are about $11\%$ and $13\%$, respectively, which indicate the effect of coupled channel plays an important role in the existence of below threshold $cc\bar{c}\bar{n}$ tetraquark state. Different from the $cc\bar{c}\bar{n}$ system, our estimations find there are no below threshold $cc\bar{c}\bar{s}$ tetraquark state with $J^P=2^+$.

In a very similar way, we can estimate the $bb\bar{b}\bar{q}$ tetraquark system, and our results are listed in Table~\ref{bbbq}. Our estimations indicate that there are no below threshold $bb\bar{b}\bar{q}$ tetraquark states. However, within the framework of QCD sum rules, the $bb\bar{b}\bar{q}$ tetraquark states with $J^{P}=0^{+}$ and  $J^{P}=1^{+}$ may be stable due to obtaining the masses below the threshold $\eta_{b}B$ and $\eta_{b}B^{\ast}$~\cite{Jiang:2017tdc}, which is different from our conclusions. It is interesting to notice that for the $cc\bar{c}\bar{n}$ system, we find one below threshold tetraquark state with $J^P=2^+$, while the mass of the corresponding state in $bb\bar{b}\bar{n}$ sector is above the threshold of $\Upsilon \bar{B}^\ast$. To find which interaction plays the dominant role in forming a below threshold $cc\bar{c}\bar{n}$ tetraquark state with $J^{P}=2^{+}$ and further check the influence of the coupled channel effect, we list the contribution of each term in the system hamiltonian in Table~\ref{part1}. As we have discussed in the above section, the potential resulting from the Goldstone-boson exchange disappeared due to the quark components of the triply heavy tetraquark system. For the $cc\bar{c}\bar{n}$ tetraquark system with $J^P=2^+$, $E_{M(J/\psi D^\ast )}$ refers to  the sum of the theoretical threshold of $J/\psi D^\ast$, which indicates the interactions between $J/\psi$ and $D^\ast$ to be zero and the system wave function is the product of the ones of $J/\psi$ and $D^\ast$. In this case, the average value of  the kinetic operator is 1800.1 MeV, and the ones of confinement and OGE terms are -1812.8 MeV and -380.5 MeV, respectively, one can obtain the threshold of $J/\psi D^\ast$ by summing over the average values of different terms and the masses of the constituent quarks. In a similar way, one can obtain the average value of the operators in the single $E_{(J/\psi D^\ast)^1}$ channel estimation, the coupled channel estimations of meson-meson configuration ($E_{cc1}$) and diquark-antidiquark configuration $E_{cc2}$, and the coupled channel estimation of both meson-meson and diquark-antidiquark configurations $E_{cc}$. For simplify, we can define the $\Delta E$ as the difference of the average values of operators between single/coupled channel cases and $E_{M(J/\psi D^\ast)}$. If the sum of $\Delta E$ for all the operators is negative, the tetraquark states are below the threshold of $J/\psi D^\ast$. From the table, one can find the sum of  $\Delta E$ for a single channel, coupled channel of each configuration is positive, while the coupled channel of both configurations is negative, which indicates the $cc\bar{c} \bar{n}$ tetraquark state with $J^P=2^+$ is a below threshold state and the coupled channel effects between different configurations are essential in forming a below threshold tetraquark state.  From the table, this result is mainly due to the strong attraction of the interaction of OGE term under the coupling of all configurations. As for $bb\bar{b}\bar{n}$ tetraquark state with $J^P=2^+$, one can find that all the $\Delta E$ is positive, which indicates the tetraquark state is above the threshold of $\Upsilon B^\ast$.


\begin{table*}[htb]
\begin{center}
\caption{\label{ccbq} The lowest-lying eigenenergies of the $cc\bar{b}\bar{n}$ $n=\{u, d\}$ and $cc\bar{b}\bar{s}$ tetraquarks in the ChQM.}
\renewcommand\arraystretch{1.35}
\begin{tabular}{p{1.25cm}<\centering p{1.35cm}<\centering p{1.5cm}<\centering p{1.35cm}<\centering p{1.35cm}<\centering p{1.35cm}<\centering p{0.5cm}<\centering p{1.5cm}<\centering p{1.35cm}<\centering p{1.5cm}<\centering p{1.35cm}<\centering p{1.35cm}<\centering p{1.35cm}<\centering}
\toprule[1pt]
\multirow{2}*{$J^{P}$} &\multirow{2}*{[i,j,k]}    &\multicolumn{4}{c}{$cc\bar{b}\bar{n}$}  & & \multicolumn{4}{c}{$cc\bar{b}\bar{s}$}   \\
\cline{3-6} \cline{8-11}
 & ~~&Channel &$E_{th}$ & $E_{sc}$  & $P(\%)$        &  &Channel &$E_{th}$ &$E_{sc}$  &$P(\%)$ \\
\cline{3-6}\cline{8-11}
\toprule[1pt]
\multirow{9}*{$0^{+}$}  &$[3,1,1]$&$( B_{c}^{+}D^{})^{1}$            &8140 &8142&$99.42\%$ &&$( B_{c}^{+}D_{s}^{+})^{1}$        &8243 &8244&$99.94\%$\\
                        &$[3,2,1]$&$(B_{c}^{\ast+}D^{\ast})^{1}$     &8307 &8309&$\sim0\%$ &&$(B_{c}^{\ast+}D_{s}^{\ast+})^{1}$ &8412 &8413&$\sim0\%$ \\
                        &$[3,1,2]$&$( B_{c}^{+}D^{})^{8}$            &     &8755&$\sim0\%$ &&$( B_{c}^{+}D_{s}^{+})^{8}$        &     &8840&$\sim0\%$  \\
                        &$[3,2,2]$&$(B_{c}^{\ast+}D^{\ast})^{8}$     &     &8756&$\sim0\%$ &&$(B_{c}^{\ast+}D_{s}^{\ast+})^{8}$ &     &8806&$\sim0\%$ \\
                        &$[4,1,3]$&$(cc)(\bar{b}\bar{n})$    &     &8735&$\sim0\%$ &&$(cc)(\bar{b}\bar{s})$             &     &8816&$\sim0\%$  \\
                        &$[4,2,4]$&$(cc)(\bar{b}\bar{n})$    &     &8657&$\sim0\%$ &&$(cc)(\bar{b}\bar{s})$             &     &8728&$\sim0\%$ \\
                        &$E_{CC1}$ &                                  &     &8142&          &&                                   &     &8244&   \\
                        &$E_{CC2}$ &                                  &     &8651&          &&                                   &     &8719&   \\
                        &$E_{CC}$ &                                   &     &8142&          &&                                   &     &8244&  \\
\midrule[1pt]
\multirow{12}*{$1^{+}$}  &$[3,3,1]$ &$(B_{c}^{+}D^{*})^{1}$        &8282 &8284&$\sim0\%$ &&$(B_{c}^{+}D_{s}^{{\ast+}})^{1}$      &8387 &8388&$\sim0\%$ \\
                         &$[3,4,1]$ &$(B_{c}^{\ast+}D^{})^{1}$     &8165 &8167&$98.00\%$ &&$(B_{c}^{\ast+}D_{s}^{+})^{1}$        &8268 &8269&$99.91\%$ \\
                         &$[3,5,1]$ &$(B_{c}^{\ast+}D^{\ast})^{1}$ &8307 &8309&$\sim0\%$ &&$(B_{c}^{\ast+}D_{s}^{\ast+})^{1}$    &8412 &8413&$\sim0\%$   \\
                         &$[3,3,2]$ &$(B_{c}^{+}D^{{\ast}})^{8}$   &     &8734&$\sim0\%$ &&$(B_{c}^{+}D_{s}^{{\ast+}})^{8}$      &     &8818&$\sim0\%$  \\
                         &$[3,4,2]$ &$(B_{c}^{\ast+}D^{})^{8}$     &     &8749&$\sim0\%$ &&$(B_{c}^{\ast+}D_{s}^{+})^{8}$        &     &8833&$\sim0\%$    \\
                         &$[3,5,2]$ &$(B_{c}^{\ast+}D^{\ast})^{8}$ &     &8739&$\sim0\%$ &&$(B_{c}^{\ast+}D_{s}^{\ast+})^{8}$    &     &8806&$\sim0\%$ \\
                         &$[4,3,3]$ &$(cc)(\bar{b}\bar{n})$&     &8724&$\sim0\%$ &&$(cc)(\bar{b}\bar{s})$                &     &8804&$\sim0\%$  \\
                         &$[4,4,4]$ &$(cc)(\bar{b}\bar{n})$&     &8644&$\sim0\%$ &&$(cc)(\bar{b}\bar{s})$                &     &8717&$\sim0\%$   \\
                         &$[4,5,4]$ &$(cc)(\bar{b}\bar{n})$&     &8662&$\sim0\%$ &&$(cc)(\bar{b}\bar{s})$                &     &8735&$\sim0\%$   \\
                         &$E_{CC1}$ &                               &     &8166&          &&                                      &     &8269&\\
                         &$E_{CC2}$ &                               &     &8640&          &&                                      &     &8710& \\
                         &$E_{CC}$  &                               &     &8166&          &&                                      &     &8269&\\
\midrule[1pt]
\multirow{6}*{$2^{+}$} &$[3,6,1]$ &$(B_{c}^{\ast+}D^{\ast})^{1}$    &8307 &8309&$99.75\%$ &&$(B_{c}^{\ast+}D_{s}^{\ast+})^{1}$ &8412 &8413&$99.96\%$ \\
                        &$[3,6,2]$ &$(B_{c}^{+}D^{\ast})^{8}$        &     &8719&$\sim0\%$ &&$(B_{c}^{+}D_{s}^{\ast+})^{8}$     &     &8818&$\sim0\%$  \\
                        &$[4,6,4]$ &$(cc)(\bar{b}\bar{n})$   &     &8671&$\sim0\%$ &&$(cc)(\bar{b}\bar{s})$             &     &8747&$\sim0\%$  \\
                        &$E_{CC1}$ &                                  &     &8309&          &&                                   &     &8413&    \\
                        &$E_{CC2}$ &                                  &     &8719&          &&                                   &     &8747&     \\
                        &$E_{CC}$ &                                   &     &8308&          &&                                   &     &8413&      \\
\bottomrule[1pt]
\end{tabular}
\end{center}
\end{table*}

\begin{table*}[!htb]
\begin{center}
\caption{\label{bbcq} The lowest-lying eigenenergies of the $bb\bar{c}\bar{n}$ $n=\{u, d\}$ and $bb\bar{c}\bar{s}$ tetraquarks in the ChQM.}
\renewcommand\arraystretch{1.35}
\begin{tabular}{p{1.25cm}<\centering p{1.35cm}<\centering p{1.5cm}<\centering p{1.35cm}<\centering p{1.35cm}<\centering p{1.35cm}<\centering p{0.5cm}<\centering p{1.5cm}<\centering p{1.35cm}<\centering p{1.5cm}<\centering p{1.35cm}<\centering p{1.35cm}<\centering p{1.35cm}<\centering}
\toprule[1pt]
\multirow{2}*{$J^{P}$} &\multirow{2}*{[i,j,k]}    &\multicolumn{4}{c}{$bb\bar{c}\bar{n}$}  &  & \multicolumn{4}{c}{$bb\bar{c}\bar{s}$}   \\
\cline{3-6} \cline{8-11}
 & ~~&Channel &$E_{th}$ & $E_{sc}$  & $P(\%)$        &  &Channel &$E_{th}$ &$E_{sc}$  &$P(\%)$ \\
\cline{3-6}\cline{8-11}
\toprule[1pt]
\multirow{9}*{$0^{+}$} &$[3,1,1]$&$(B_{c}^{-}\bar{B})^{1}$             &11554&11557 &$94.25\%$ &&$(B_{c}^{-}\bar{B}_{s})^{1}$            &11642&11643 &99.75\%\\
                       &$[3,2,1]$&$(B_{c}^{\ast-}\bar{B}^{\ast})^{1}$  &11625&11627 &$1.45\%$  &&$(B_{c}^{\ast-}\bar{B}_{s}^{\ast})^{1}$ &11715&11716 &$\sim0\%$ \\
                       &$[3,1,2]$&$(B_{c}^{-}\bar{B})^{8}$             &     &12003 &$\sim0\%$ &&$(B_{c}^{-}\bar{B}_{s})^{8}$            &     &12062 &$\sim0\%$  \\
                       &$[3,2,2]$&$(B_{c}^{\ast-}\bar{B}^{\ast})^{8}$  &     &12082 &$1.24\%$  &&$(B_{c}^{\ast-}\bar{B_{s}}^{\ast})^{8}$ &     &12114 &$\sim0\%$ \\
                       &$[4,1,3]$&$(bb)(\bar{c}\bar{n})$      &     &12146 &$\sim0\%$ &&$(bb)(\bar{c}\bar{s})$                  &     &12193 &$\sim0\%$  \\
                       &$[4,2,4]$&$(bb)(\bar{c}\bar{n})$      &     &11827 &$2.31$\% &&$(bb)(\bar{c}\bar{s})$                   &     &11893 &$\sim0\%$  \\
                       &$E_{CC1}$&                                     &     &11556 &          &&                                        &     &11643 &   \\
                       &$E_{CC2}$&                                     &     &11826 &          &&                                        &     &11893 &   \\
                       &$E_{CC}$ &                                     &     &11552 &          &&                                        &     &11643 &  \\
\midrule[1pt]
\multirow{12}*{$1^{+}$} &$[3,3,1]$ &$(B_{c}^{-}\bar{B^{\ast}})^{1}$     &11600&11602  &15.71\%  &&$(B_{c}^{-}\bar{B_{s}^{\ast}})^{1}$     &11690&11691  &$\sim0\%$  \\
                        &$[3,4,1]$ &$(B_{c}^{\ast-}\bar{B})^{1}$        &11579&11582  &58.26\%  &&$(B_{c}^{\ast-}\bar{B_{s}})^{1}$        &11667&11668  &98.59\%  \\
                        &$[3,5,1]$ &$(B_{c}^{\ast-}\bar{B}^{\ast})^{1}$ &11625&11627  &$\sim0\%$&&$(B_{c}^{\ast-}\bar{B_{s}}^{\ast})^{1}$ &11715&11716  &$\sim0\%$   \\
                        &$[3,3,2]$ &$(B_{c}^{-}\bar{B^{\ast}})^{8}$     &     &11987  &$\sim0\%$&&$(B_{c}^{-}\bar{B_{s}^{\ast}})^{8}$     &     &12043  &$\sim0\%$  \\
                        &$[3,4,2]$ &$(B_{c}^{\ast-}\bar{B})^{8}$        &     &11990  &$\sim0\%$&&$(B_{c}^{\ast-}\bar{B_{s}})^{8}$        &     &12046  &$\sim0\%$   \\
                        &$[3,5,2]$ &$(B_{c}^{\ast-}\bar{B}^{\ast})^{8}$ &     &12010  &6.16\%   &&$(B_{c}^{\ast-}\bar{B_{s}}^{\ast})^{8}$ &     &12050  &$\sim0\%$   \\
                        &$[4,3,3]$ &$(bb)(\bar{c}\bar{n})$     &     &12110  &1.58\%   &&$(bb)(\bar{c}\bar{s})$                  &     &12157  &$\sim0\%$   \\
                        &$[4,4,4]$ &$(bb)(\bar{c}\bar{n})$     &     &11759  &16.57\%  &&$(bb)(\bar{c}\bar{s})$                  &     &11827  &$\sim0\%$   \\
                        &$[4,5,4]$ &$(bb)(\bar{c}\bar{n})$     &     &11828  &$\sim0\%$&&$(bb)(\bar{c}\bar{s})$                  &     &11896  &$\sim0\%$   \\
                        &$E_{CC1}$ &                                    &     &11581  &         &&                                        &     &11668  &\\
                        &$E_{CC2}$ &                                    &     &11759  &         &&                                        &     &11826  & \\
                        &$E_{CC}$ &                                     &     &11566  &         &&                                        &     &11668  &\\
\midrule[1pt]
\multirow{6}*{$2^{+}$} &$[3,6,1]$ &$(B_{c}^{\ast-}\bar{B}^{\ast})^{1}$ &11625&11627  &68.00\%  &&$(B_{c}^{\ast-}\bar{B_{s}}^{\ast})^{1}$ &11715&11716  &99.57\% \\
                        &$[3,6,2]$ &$(B_{c}^{-}\bar{B}^{\ast})^{8}$     &     &11961  &9.79\%   &&$(B_{c}^{-}\bar{B_{s}}^{\ast})^{8}$     &     &12031  &$\sim0\%$  \\
                        &$[4,6,4]$ &$(bb)(\bar{c}\bar{n})$     &     &11833  &22.21\%  &&$(bb)(\bar{c}\bar{s})$                  &     &11901  &$\sim0\%$ \\
                        &$E_{CC1}$ &                                    &     &11626  &         &&                                        &     &11716  &    \\
                        &$E_{CC2}$ &                                    &     &11833  &         &&                                        &     &11901  &     \\
                        &$E_{CC}$  &                                    &     &11613  &         &&                                        &     &11716  &      \\
\bottomrule[1pt]
\end{tabular}
\end{center}
\end{table*}

\begin{table*}[ht]
\begin{center}
\caption{\label{part2}
The same as Table \ref{part1} but for $cc\bar{b} \bar{n}$ and $bb\bar{c}\bar{n}$ tetraquark states with $J^P=0^+$, $1^+$ and $2^+$. }
\renewcommand\arraystretch{1.5}
  \begin{tabular}{p{2.cm}<\centering p{1.5cm}<\centering p{1.5cm}<\centering p{1.5cm}<\centering p{1.5cm}<\centering <\centering p{0.5cm}<\centering p{1.2cm}<\centering p{1.5cm}<\centering p{1.5cm}<\centering p{1.5cm}<\centering}
 \toprule[1pt]
     & & $\langle H_{T}\rangle$ & $\langle V_{CON}\rangle$ & $\langle V_{OGE}\rangle $  & & & $\langle H_{T}\rangle $ & $\langle V_{CON}\rangle $ & $\langle V_{OGE} \rangle$ \\
    \cline{2-5} \cline{7-10}
 \multirow{9}*{$J^{P}=0^{+}$} &$E_{(B_{c}^{+}D)^{1}}$          &1662.7 &-1984.5 &-416.5 &&$E_{(B_{c}^{-}\bar{B})^{1}}$ &1522.7 &-1880.4 &-349.9 \\
   &$E_{CC1}$                                                  &1662.4 &-1985.2 &-416.2 &&$E_{CC1}$                    &1522.4 &-1882.7 &-348.9  \\
   &$E_{CC2}$                                                  &1861.4 &-1751.3 &-339.7 &&$E_{CC2}$                    &1698.3 &-1550.5 &-586.6  \\
   &$E_{CC}$                                                   &1660.3 &-1984.9 &-414.6 &&$E_{CC}$                     &1521.2 &-1882.8 &-351.3  \\
   &$E_{M(B_{c}^{+}D)}$                                  &1660.2 &-1984.6 &-416.6 &&$E_{M(B_{c}^{-}\bar{B})}$    &1519.3 &-1880.3 &-350.0  \\
   &$\Delta E_{(B_{c}^{+}D)^{1}}$                              &2.5    &0.1     &0.1    &&$\Delta E_{(B_{c}^{-}\bar{B})^{1}}$  &3.5    &-0.1     &0.1\\
   &$\Delta  E_{CC1}$                                          &2.2    &-0.6    &0.4    &&$\Delta  E_{CC1}$                &3.1    &-2.4     &1.1\\
   &$\Delta  E_{CC2}$                                          &201.2  &233.3   &76.9   &&$\Delta  E_{CC2}$                &179.0  &329.8    &-236.6 \\
   &$\Delta  E_{CC}$                                           &0.1    &-0.3    &2.0    &&$\Delta  E_{CC}$                 &2.0    &-2.8     &-1.3\\
    \midrule[1pt]
  \multirow{9}*{$J^{P}=1^{+}$} &$E_{(B_{c}^{\ast+}D)^{1}}$     &1662.7 &-1984.5 &-391.5 &&$E_{(B_{c}^{\ast-}\bar{B})^{1}}$  &1522.7 &-1880.4 &-324.9 \\
   &$E_{CC1}$                                                  &1662.3 &-1985.3 &-391.2 &&$E_{CC1}$                         &1521.9 &-1883.5 &-322.8  \\
   &$E_{CC2}$                                                  &1851.2 &-1741.1 &-351.0 &&$E_{CC2}$                         &1695.4 &-1548.2 &-652.8  \\
   &$E_{CC}$                                                   &1662.4 &-1985.3 &-391.2 &&$E_{CC}$                          &1520.3 &-1884.9 &-334.5  \\
   &$E_{M(B_{c}^{\ast+}D)}$                                    &1660.2 &-1984.5 &-391.6 &&$E_{M(B_{c}^{\ast-}\bar{B})}$     &1519.3 &-1880.3 &-325.0  \\
   &$\Delta E_{(B_{c}^{\ast+}D)^{1}}$                          &2.5    &0.0     &0.1    &&$\Delta E_{(B_{c}^{\ast-}\bar{B})^{1}}$  &3.4    &-0.1     &0.1\\
   &$\Delta  E_{CC1}$                                          &2.1    &-0.8    &0.4    &&$\Delta  E_{CC1}$                        &2.6    &-3.2     &2.2\\
   &$\Delta  E_{CC2}$                                          &197.0  &243.4   &40.6   &&$\Delta  E_{CC2}$                        &176.1  &332.1    &-327.8 \\
   &$\Delta  E_{CC}$                                           &2.2    &-0.8    &0.4    &&$\Delta  E_{CC}$                         &1.0    &-4.6    &-9.5\\
   \midrule[1pt]
   \multirow{9}*{$J^{P}=2^{+}$}&$E_{(B_{c}^{\ast+}D^{\ast})^{1}}$&1662.7 &-1984.5 &-249.5  &&$E_{(B_{c}^{\ast-}\bar{B}^{\ast})^{1}}$&1522.7 &-1880.4 &-279.9 \\
   &$E_{CC1}$                                                   &1662.7  &-1985.1 &-249.4 &&$E_{CC1}$                              &1522.7 &-1881.9 &-279.8  \\
   &$E_{CC2}$                                                   &1839.2  &-1728.8 &-319.9 &&$E_{CC2}$                              &1694.0 &-1547.0 &-579.4  \\
   &$E_{CC}$                                                    &1661.5  &-1983.8 &-249.9 &&$E_{CC}$                               &1521.8 &-1883.5 &-289.7  \\
   &$E_{M(B_{c}^{\ast+}D^{\ast})}$                              &1660.2  &-1984.5 &-249.6 &&$E_{M(B_{c}^{\ast-}\bar{B}^{\ast})}$   &1520.4 &-1880.3 &-280.0  \\
   &$\Delta E_{(B_{c}^{\ast+}D^{\ast})^{1}}$                    &2.5     &0.0     &0.1    &&$\Delta  E_{(B_{c}^{\ast-}\bar{B}^{\ast})^{1}}$     &2.3    &-0.1     &0.1\\
   &$\Delta  E_{CC1}$                                           &2.5     &0.4     &0.2    &&$\Delta  E_{CC1}$                                   &2.3    &-1.6     &0.2\\
   &$\Delta  E_{CC2}$                                           &179.0   &255.7   &-70.4  &&$\Delta  E_{CC2}$                                   &173.6  &333.3    &-299.4 \\
   &$\Delta  E_{CC}$                                            &1.3     &0.7     &-0.3   &&$\Delta  E_{CC}$                                    &1.2    &-3.2     &-9.7\\

  \bottomrule[1pt]
\end{tabular}
\end{center}
\end{table*}


\subsection{The $QQ\bar{Q^{\prime}}\bar{q}$ system}

In Table \ref{ccbq}, we present our estimations of the eigenenergies of the $cc\bar{b}\bar{q}$ system with $J^P=0^+,\ 1^+$ and $2^+$, respectively. For the case of $cc\bar{b}\bar{n}$ tetraquark with $J^P=0^+$, we find there are four meson-meson channels and two diquark-antidiquark channels. The lowest physical threshold of $cc\bar{b}\bar{n}$ is the one of $B_c^+ D$, which is $8140$ MeV. The eigenenergies obtained from the single channel, coupled channel in each configuration, and the full coupled channel estimations are all above the threshold of $B_c^+ D$. From the full coupled channel estimations one can find the dominant component of $cc\bar{b}\bar{n}$ tetraquark state with $J^P=0^+$ is $B_c^+ D$. As for the $cc\bar{b}\bar{n}$ tetraquark states with $J^P=1^+$, there are six meson-meson and three diquark-antidiquark channels, respectively. The lowest physical threshold is the one of $B_c^+ D^\ast$, which is 8282 MeV. Similar to the case of $0^+$, the eigenenergies obtained from the single channel, coupled channel in each configuration, and the full coupled channel estimations are all above the threshold of $B_c^+ D^\ast$. Similarly, there are two meson-meson and one diquark-antidiquark channels in the $cc\bar{b}\bar{n}$ tetraquark system with $J^P=2^+$, and our estimations also indicate that there is no below threshold $cc\bar{b}\bar{n}$ tetraquark state with $J^P=2^+$. Similarly, we can analyze the $cc\bar{b}\bar{s}$ tetraquark system, and we find all the eigenenergies of the $cc\bar{b}\bar{s}$ tetraquark are above the lowest thresholds of the corresponding physical channels.

As for the $bb\bar{c}\bar{q}$ tetraquark system, the estimated eigenenergies are listed in Table~\ref{bbcq}. For the $bb\bar{c}\bar{n}$ tetraquark states with $J^P=0^+$, we find the lowest threshold of physical channel is the one of $B_c^- \bar{B}$, which is 11554 MeV. The eigenenergies obtained from the single channel estimations and coupled channel estimations in each configuration are above the threshold of $B_c^- \bar{B}$. While considering the coupled channel effects of meson-meson and diquark-antidiquark configurations simultaneously, we find the eigenenergies of $bb\bar{c}\bar{n}$ tetraquark with $J^P=0^+$ is 11552 MeV, which is about 2 MeV below the threshold of $B_c^- \bar{B}$. In this tetraquark state, the dominant component is $B_c^- \bar{B}$ and its percentage is about 94.25. As for the $bb\bar{c}\bar{n}$ tetraquark states with $J^P=1^+$, the lowest physical channel is $B_c^- \bar{B}^\ast$ with the threshold to be 11579 MeV. We find that the eigenenergies obtained from single channel estimations and coupled channel estimations in each configuration are all above the threshold of $B_c^- \bar{B}^\ast$, while the full coupled channel estimations indicate that the eigenenergy of the $bb\bar{c}\bar{n}$ tetraquark states with $J^P=1^+$ is 11566 MeV, which is about 13 MeV below the threshold of $B_c^- \bar{B}^\ast$.  In this tetraquark state, the dominant component is $B_c^- \bar{B}^\ast$ and its percentage is about 58.26, while the $(B_c^- \bar{B}^\ast)^1$ and $(B_c^- \bar{B}^\ast)^8$ channels in the meson-meson configuration and $(bb)(\bar{c}\bar{n})$ channel with $[i,j,k]=[4,4,4]$ in the diquark-antidiquark configuration are also important with the percentage to be $15.71$, 6.16 and 16.57, respectively. For the $J^P=2^+$ case, there is only one physical channel for $bb\bar{c}\bar{n}$ tetraquark state, which is $B_c^{\ast -} \bar
{B}^\ast$ with the threshold to be 11625 MeV. Similar to the case of $0^+ $ and $2^+$, the eigenenergies obtained from the single channel estimations and the coupled channel estimations in each configuration are all above the threshold of $B_c^{\ast-} \bar{B}^{\ast}$. When we consider both meson-meson and diquark-antidiquark configurations simultaneously, the eigenenergy of $bb\bar{c}\bar{n}$ tetraquark state with $J^P=2^+$ is estimated to be 11613 MeV, which is about 12 MeV below the threshold of $B_c^{\ast-} \bar{B}^{\ast}$ and the percentage of different channels are 68.00, 9.79 and 22.21 for $(B_c^{\ast-} \bar{B}^{\ast})^1$, $(B_c^{\ast-} \bar{B}^{\ast})^8$ channels and $(bb)(\bar{c}\bar{n})$ channel with $[i,j,k]=[4,6,4]$, respectively. Different from the $bb\bar{c} \bar{n}$ tetraquark system, our estimations indicate the eigenenergies of $bb\bar{c} \bar{s}$ tetraquark states with different $J^P$ quantum numbers are all above the lowest threshold of the corresponding physical channels.

From our estimations, we find there is no below threshold $QQ\bar{Q}^\prime \bar{s}$ tetraquark state. But for $QQ\bar{Q}^\prime \bar{n}$ tetraquark system, we find the eigenenergies of all the $S$-wave ground $bb\bar{c}\bar{n}$ tetraquark states with different $J^P$ quantum numbers are below the lowest threshold of the corresponding physical channels, which is much different with $cc\bar{b}\bar{n}$ case. To further compare the spectrum of $cc\bar{b}\bar{n}$ and $bb\bar{c} \bar{n}$, we list the average values of each operator in the Hamiltonian of the tetraquark systems in Table \ref{part2}. It is interesting to notice that in the full coupled channel estimation all the eigenenergies of the $bb\bar{c}\bar{n}$ tetraquark states are below the corresponding lowest physical threshold, while the eigenenergies of the $cc\bar{b}\bar{n}$ are all above the corresponding lowest physical threshold. By comparing the average values of the operators in the Hamiltonian of the $cc\bar{b}\bar{n}$ and $bb\bar{c}\bar{n}$ tetraquark system, one finds the dominant difference is the average values of $V_{OGE}$, especially in the case of coupled channel estimations in the diquark-antidiquark configurations. The average values of $V_{OGE}$ are negative, which indicates that the OGE potential is attractive. However, for the $cc\bar{b}\bar{n}$ tetraquark states with $J^P=0^+$ and $2^+$, the attractions become weak when we consider coupled channel effects in each configuration, and for $J^P=2^+$ case, the attraction becomes stronger in the diquark-antidiquark coupled channel estimations. For the $bb\bar{c}\bar{n}$ tetraquark states, we find that the attractions become much stronger in the diquark-antidiquark coupled channel estimations, although the attractions caused by the confinement potential become weak and the eigenenergies obtained in the diquark-antidiquark coupled channel estimations are still above the corresponding lowest physical threshold. But when we consider the coupled channel effects in both configurations, the eigenenergies of the $bb\bar{c}\bar{n}$ are below the corresponding lowest threshold of the physical channels.

\subsection{The $QQ^{\prime}\bar{Q}\bar{q}$ system}

\begin{table*}[htb]
\begin{center}
\caption{\label{cbcq} The lowest-lying eigenenergies of the $bc\bar{c}\bar{n}$ $n=\{u, d\}$ and $bc\bar{c}\bar{s}$ tetraquarks in the ChQM.}
\renewcommand\arraystretch{1.25}
\begin{tabular}{p{1.25cm}<\centering p{1.35cm}<\centering p{1.5cm}<\centering p{1.35cm}<\centering p{1.35cm}<\centering p{1.35cm}<\centering p{0.5cm}<\centering p{1.5cm}<\centering p{1.35cm}<\centering p{1.5cm}<\centering p{1.35cm}<\centering p{1.35cm}<\centering p{1.35cm}<\centering}
\toprule[1pt]
\multirow{2}*{$J^{P}$} &\multirow{2}*{[i,j,k]}    &\multicolumn{4}{c}{$bc\bar{c}\bar{n}$}   & \multicolumn{4}{c}{$bc\bar{c}\bar{s}$}   \\
\cline{3-6} \cline{8-11}
 & ~~&Channel &$E_{th}$ & $E_{sc}$  & $P(\%)$        &  &Channel &$E_{th}$ &$E_{sc}$  &$P(\%)$ \\
\cline{3-6}\cline{8-11}
\toprule[1pt]
\multirow{15}*{$0^{+}$} &$[5,1,1]$&$(\eta_{c}\bar{B})^{1}$       &8266 &8266&$\sim0\%$ &&$(\eta_{c}\bar{B_{s}})^{1}$       &8351 &8352&$\sim0\%$\\
                       &$[5,2,1]$&$(J/\psi\bar{B^{\ast}})^{1}$   &8422 &8424&$\sim0\%$ &&$(J/\psi \bar{B_{s}^{\ast}})^{1}$ &8512 &8513&$\sim0\%$  \\
                       &$[5,1,2]$&$(\eta_{c}\bar{B})^{8}$        &     &8717&$\sim0\%$ &&$(\eta_{c}\bar{B_{s}})^{8}$       &     &8796&$\sim0\%$  \\
                       &$[5,2,2]$&$(J/\psi\bar{B^{\ast}})^{8}$   &     &8663&$\sim0\%$ &&$(J/\psi\bar{B_{s}^{\ast}})^{8}$  &     &8733&$\sim0\%$ \\
                       &$[6,1,1]$&$(D^{}B_{c}^{-})^{1}$          &8140 &8142&$98.94\%$ &&$(D_{s}^{+}B_{c}^{-})^{1}$        &8243 &8244&$99.92\%$  \\
                       &$[6,2,1]$&$(D^{\ast}B_{c}^{\ast-})^{1}$  &8307 &8309&$\sim0\%$ &&$(D_{s}^{\ast+}B_{c}^{\ast-})^{1}$&8412 &8413&$\sim0\%$   \\
                       &$[6,1,2]$&$(D^{}B_{c}^{-})^{8}$          &     &8870&$\sim0\%$ &&$(D_{s}^{+}B_{c}^{-})^{8}$        &     &8924&$\sim0\%$  \\
                       &$[6,2,2]$&$(D^{\ast}B_{c}^{\ast-})^{8}$  &     &8815&$\sim0\%$ &&$(D_{s}^{\ast+}B_{c}^{\ast-})^{8}$&     &8860&$\sim0\%$   \\
                       &$[7,1,3]$&$(bc)(\bar{c}\bar{n})$&     &8864&$\sim0\%$ &&$(bc)(\bar{c}\bar{s})$            &     &8924&$\sim0\%$ \\
                       &$[7,1,4]$&$(bc)(\bar{c}\bar{n})$&     &8648&$\sim0\%$ &&$(bc)(\bar{c}\bar{s})$            &     &8722&$\sim0\%$  \\
                       &$[7,2,3]$&$(bc)(\bar{c}\bar{n})$&     &8864&$\sim0\%$ &&$(bc)(\bar{c}\bar{s})$            &     &8924&$\sim0\%$  \\
                       &$[7,2,4]$&$(bc)(\bar{c}\bar{n})$&     &8643&$\sim0\%$ &&$(bc)(\bar{c}\bar{s})$            &     &8715&$\sim0\%$  \\
                       &$E_{CC1}$ &                              &     &8141&          &&                                  &     &8244 &   \\
                       &$E_{CC2}$ &                              &     &8647&          &&                                  &     &8711 &   \\
                       &$E_{CC}$ &                               &     &8141&          &&                                  &     &8244 &  \\
\midrule[1pt]
\multirow{21}*{$1^{+}$} &$[5,3,1]$ &$(\eta_{c}\bar{B^{\ast}})^{1}$ &8309 &8311&$\sim0\%$ &&$(\eta_{c}\bar{B_{s}^{\ast}})^{1}$ &8399 &8400&$\sim0\%$   \\
                        &$[5,4,1]$ &$(J/\psi\bar{B})^{1}$          &8377 &8379&$\sim0\%$ &&$(J/\psi\bar{B_{s}})^{1}$          &8464 &8465&$\sim0\%$  \\
                        &$[5,5,1]$ &$(J/\psi\bar{B^{\ast}})^{1}$   &8422 &8424&$\sim0\%$ &&$(J/\psi\bar{B_{s}^{\ast}})^{1}$   &8512 &8513&$\sim0\%$   \\
                        &$[5,3,2]$ &$(\eta_{c}\bar{B^{\ast}})^{8}$ &     &8711&$\sim0\%$ &&$(\eta_{c}\bar{B_{s}^{\ast}})^{8}$ &     &8790&$\sim0\%$   \\
                        &$[5,4,2]$ &$(J/\psi\bar{B})^{8}$          &     &8702&$\sim0\%$ &&$(J/\psi\bar{B_{s}})^{8}$          &     &8781&$\sim0\%$   \\
                        &$[5,5,2]$ &$(J/\psi\bar{B^{\ast}})^{8}$   &     &8680&$1.77\%$  &&$(J/\psi\bar{B_{s}^{\ast}})^{8}$   &     &8754&$\sim0\%$  \\
                        &$[6,3,1]$ &$(D^{}B_{c}^{\ast-})^{1}$      &8165 &8167&$91.57\%$ &&$(D_{s}^{+}B_{c}^{\ast-})^{1}$     &8268 &8269&$99.76\%$    \\
                        &$[6,4,1]$ &$(D^{\ast}B_{c}^{-})^{1}$      &8282 &8284&$\sim0\%$ &&$(D_{s}^{\ast+}B_{c}^{-})^{1}$     &8387 &8388&$\sim0\%$    \\
                        &$[6,5,1]$ &$(D^{\ast}B_{c}^{\ast-})^{1}$  &8307 &8309&$\sim0\%$ &&$(D_{s}^{\ast+}B_{c}^{\ast-})^{1}$ &8412 &8413&$\sim0\%$    \\
                        &$[6,3,2]$ &$(D^{}B_{c}^{\ast-})^{8}$      &     &8867&$\sim0\%$ &&$(D_{s}^{+}B_{c}^{\ast-})^{8}$     &     &8921&$\sim0\%$  \\
                        &$[6,4,2]$ &$(D^{\ast}B_{c}^{-})^{8}$      &     &8853&$\sim0\%$ &&$(D_{s}^{\ast+}B_{c}^{-})^{8}$     &     &8906&$\sim0\%$     \\
                        &$[6,5,2]$ &$(D^{\ast}B_{c}^{\ast-})^{8}$  &     &8833&$\sim0\%$ &&$(D_{s}^{\ast+}B_{c}^{\ast-})^{8}$ &     &8882&$\sim0\%$   \\
                        &$[7,3,3]$ &$(bc)(\bar{c}\bar{n})$&     &8828&$\sim0\%$ &&$(bc)(\bar{c}\bar{s})$             &     &8888&$\sim0\%$   \\
                        &$[7,3,4]$ &$(bc)(\bar{c}\bar{n})$&     &8640&$\sim0\%$ &&$(bc)(\bar{c}\bar{s})$             &     &8715&$\sim0\%$   \\
                        &$[7,4,3]$ &$(bc)(\bar{c}\bar{n})$&     &8858&$\sim0\%$ &&$(bc)(\bar{c}\bar{s})$             &     &8918&$\sim0\%$    \\
                        &$[7,4,4]$ &$(bc)(\bar{c}\bar{n})$&     &8582&$1.17\%$  &&$(bc)(\bar{c}\bar{s})$             &     &8656&$\sim0\%$   \\
                        &$[7,5,3]$ &$(bc)(\bar{c}\bar{n})$&     &8803&$\sim0\%$ &&$(bc)(\bar{c}\bar{s})$             &     &8722&$\sim0\%$  \\
                        &$[7,5,4]$ &$(bc)(\bar{c}\bar{n})$&     &8648&$1.19\%$  &&$(bc)(\bar{c}\bar{s})$             &     &8858&$\sim0\%$  \\
                        &$E_{CC1}$ &                               &     &8167&          &&                                   &     &8269  &\\
                        &$E_{CC2}$ &                               &     &8554&          &&                                   &     &8633  & \\
                        &$E_{CC}$ &                                &     &8159&          &&                                   &     &8269  &\\
\midrule[1pt]
\multirow{9}*{$2^{+}$} &$[5,6,1]$ &$(J/\psi\bar{B^{\ast}})^{1}$      &8422 &8424&$\sim0\%$&&$(J/\psi\bar{B_{s}^{\ast}})^{1}$      &8512 &8513&$\sim0\%$  \\
                       &$[6,6,1]$ &$(J/\psi\bar{B^{\ast}})^{8}$      &     &8713&$11.04\%$&&$(J/\psi\bar{B_{s}^{\ast}})^{8}$      &     &8796&$1.59\%$ \\
                       &$[5,6,2]$ &$(D^{\ast}B_{c}^{\ast-})^{1}$     &8307 &8309&$72.10\%$&&$(D_{s}^{\ast+}B_{c}^{\ast-})^{1}$     &8412 &8413&$95.29\%$  \\
                       &$[6,6,2]$ &$(D^{\ast}B_{c}^{\ast-})^{8}$     &     &8866&$1.38\%$ &&$(D_{s}^{\ast+}B_{c}^{\ast-})^{8}$     &     &8924&$\sim0\%$  \\
                       &$[7,6,3]$ &$(bc)(\bar{c}\bar{n})$   &     &8841&$4.53\%$ &&$(bc)(\bar{c}\bar{s})$                &     &8906&$\sim0\%$  \\
                       &$[7,6,4]$ &$(bc)(\bar{c}\bar{n})$   &     &8657&$10.77\%$&&$(bc)(\bar{c}\bar{s})$                &     &8734&$1.94\%$  \\
                       &$E_{CC1}$ &                                  &     &8308&          &&                                     &     &8413&    \\
                       &$E_{CC2}$ &                                  &     &8623&          &&                                     &     &8708&     \\
                       &$E_{CC}$  &                                  &     &8273&          &&                                     &     &8410&      \\
\bottomrule[1pt]
\end{tabular}
\end{center}
\end{table*}

\begin{table*}[htb]
\begin{center}
\caption{\label{bcbq} The lowest-lying eigenenergies of the $bc\bar{b}\bar{n}$ $n=\{u, d\}$ and $bc\bar{b}\bar{s}$ tetraquarks in the ChQM.}
\renewcommand\arraystretch{1.30}
\begin{tabular}{p{1.25cm}<\centering p{1.35cm}<\centering p{1.5cm}<\centering p{1.35cm}<\centering p{1.35cm}<\centering p{1.35cm}<\centering p{0.5cm}<\centering p{1.5cm}<\centering p{1.35cm}<\centering p{1.5cm}<\centering p{1.35cm}<\centering p{1.35cm}<\centering p{1.35cm}<\centering}
\toprule[1pt]
\multirow{2}*{$J^{P}$} &\multirow{2}*{[i,j,k]}    &\multicolumn{4}{c}{$bc\bar{b}\bar{n}$}   & \multicolumn{4}{c}{$bc\bar{b}\bar{s}$}   \\
\cline{3-6} \cline{8-11}
 & ~~&Channel &$E_{th}$ & $E_{sc}$  & $P(\%)$        &  &Channel &$E_{th}$ &$E_{sc}$  &$P(\%)$ \\
\cline{3-6}\cline{8-11}
\toprule[1pt]
\multirow{15}*{$0^{+}$} &$[5,1,1]$&$(\eta_{b}D^{})^{1}$                &11264&11266 &$99.99\%$  &&$(\eta_{b}D_{s}^{+})^{1}$               &11367&11368 &$99.99\%$\\
                        &$[5,2,1]$&$(\Upsilon D^{\ast})^{1}$           &11467&11469 &$\sim0\%$  &&$(\Upsilon D_{s}^{\ast+})^{1}$          &11572&11573 &$\sim0\%$  \\
                        &$[5,1,2]$&$(\eta_{b}D^{})^{8}$                &     &12085 &$\sim0\%$  &&$(\eta_{b}D_{s}^{+})^{1})^{8}$          &     &12135 &$\sim0\%$   \\
                        &$[5,2,2]$&$(\Upsilon D^{\ast})^{8}$           &     &12050 &$\sim0\%$  &&$(\Upsilon D_{s}^{\ast+})^{8}$          &     &12094 &$\sim0\%$  \\
                        &$[6,1,1]$&$(\bar{B}B_{c}^{+})^{1}$                  &11555&11557 &$\sim0\%$  &&$(\bar{B_{s}}B_{c}^{+})^{1}$                  &11642&11643 &$\sim0\%$   \\
                        &$[6,2,1]$&$(\bar{B^{\ast}}B_{c}^{\ast+})^{1}$       &11625&11627 &$\sim0$\%  &&$(\bar{B_{s}^{\ast}}B_{c}^{\ast+})^{1}$       &11715&11716 &$\sim0\%$   \\
                        &$[6,1,2]$&$(\bar{B}B_{c}^{+})^{8}$            &     &12002 &$\sim0\%$  &&$(\bar{B_{s}}B_{c}^{+})^{8}$            &     &12035 &$\sim0\%$  \\
                        &$[6,2,2]$&$(\bar{B^{\ast}}B_{c}^{\ast+})^{8}$ &     &11957 &$\sim0$\%  &&$(\bar{B_{s}^{\ast}}B_{c}^{\ast+})^{8}$ &     &11982 &$\sim0\%$   \\
                        &$[7,1,3]$&$(bc)(\bar{b}\bar{n})$              &     &12039 &$\sim0\%$  &&$(bc)(\bar{b}\bar{s})$                  &     &12077 &$\sim0\%$  \\
                        &$[7,1,4]$&$(bc)(\bar{b}\bar{n})$              &     &11899 &$\sim0\%$  &&$(bc)(\bar{b}\bar{s})$                  &     &11951 &$\sim0\%$   \\
                        &$[7,2,3]$&$(bc)(\bar{b}\bar{n})$              &     &11989 &$\sim0\%$  &&$(bc)(\bar{b}\bar{s})$                  &     &12018 &$\sim0\%$   \\
                        &$[7,2,4]$&$(bc)(\bar{b}\bar{n})$              &     &11925 &$\sim0\%$  &&$(bc)(\bar{b}\bar{s})$                  &     &11976 &$\sim0\%$   \\
                        &$E_{CC1}$ &                                   &     &11266 &           &&                                        &     &11368 &   \\
                        &$E_{CC2}$ &                                   &     &11879 &           &&                                        &     &11912 &   \\
                        &$E_{CC}$ &                                    &     &11266 &           &&                                        &     &11368 &  \\
\midrule[1pt]
\multirow{21}*{$1^{+}$} &$[5,3,1]$ &$(\eta_{b}D^{\ast})^{1}$          &11406&11408  &$\sim0\%$  &&$(\eta_{b}D_{s}^{+\ast})^{1}$         &11511&11512  &$\sim0\%$   \\
                        &$[5,4,1]$ &$(\Upsilon D^{})^{1}$             &11325&11327  &99.99\%    &&$(\Upsilon D_{s}^{+})^{1}$            &11428&11429  &99.99\%  \\
                        &$[5,5,1]$ &$(\Upsilon D^{\ast})^{1}$         &11467&11433  &$\sim0\%$  &&$(\Upsilon D_{s}^{+\ast})^{1}$        &11572&11573  &$\sim0\%$   \\
                        &$[5,3,2]$ &$(\eta_{b}D^{\ast})^{8}$          &     &12067  &$\sim0\%$  &&$(\eta_{b}D_{s}^{+\ast})^{8}$         &     &12117  &$\sim0\%$   \\
                        &$[5,4,2]$ &$(\Upsilon D^{})^{8}$             &     &12082  &$\sim0\%$  &&$(\Upsilon D_{s}^{+})^{8}$            &     &12131  &$\sim0\%$   \\
                        &$[5,5,2]$ &$(\Upsilon D^{\ast})^{8}$         &     &12057  &$\sim0\%$  &&$(\Upsilon D_{s}^{+\ast})^{8}$        &     &12104  &$\sim0\%$   \\
                        &$[6,3,1]$ &$(\bar{B}B_{c}^{\ast+})^{1}$        &11580&11582  &$\sim0\%$  &&$(\bar{B_{s}}B_{c}^{\ast+})^{1}$       &11667&11668  &$\sim0\%$    \\
                        &$[6,4,1]$ &$(\bar{B^{\ast}}B_{c}^{+})^{1}$     &11600&11602  &$\sim0\%$  &&$(\bar{B_{s}^{\ast}}B_{c}^{+})^{1}$    &11690&11691  &$\sim0\%$   \\
                        &$[6,5,1]$ &$(\bar{B^{\ast}}B_{c}^{\ast+})^{1}$ &11625&11627  &$\sim0\%$  &&$(\bar{B_{s}^{\ast}}B_{c}^{\ast+})^{1}$&11715&11716  &$\sim0\%$    \\
                        &$[6,3,2]$ &$(\bar{B}B_{c}^{\ast+})^{8}$        &     &11999  &$\sim0\%$  &&$(\bar{B_{s}}B_{c}^{\ast+})^{8}$       &     &12032  &$\sim0\%$    \\
                        &$[6,4,2]$ &$(\bar{B^{\ast}}B_{c}^{+})^{8}$     &     &11997  &$\sim0\%$  &&$(\bar{B_{s}^{\ast}}B_{c}^{+})^{8}$    &     &12029  &$\sim0\%$    \\
                        &$[6,5,2]$ &$(\bar{B^{\ast}}B_{c}^{\ast+})^{8}$ &     &11975  &$\sim0\%$  &&$(\bar{B_{s}^{\ast}}B_{c}^{\ast+})^{8}$&     &12004  &$\sim0\%$    \\
                        &$[7,3,3]$ &$(bc)(\bar{b}\bar{n})$    &     &12027  &$\sim0\%$  &&$(bc)(\bar{b}\bar{s})$                &     &12065  &$\sim0\%$   \\
                        &$[7,3,4]$ &$(bc)(\bar{b}\bar{n})$    &     &11921  &$\sim0\%$  &&$(bc)(\bar{b}\bar{s})$                &     &11975  &$\sim0\%$    \\
                        &$[7,4,3]$ &$(bc)(\bar{b}\bar{n})$    &     &12032  &$\sim0\%$  &&$(bc)(\bar{b}\bar{s})$                &     &12070  &$\sim0\%$   \\
                        &$[7,4,3]$ &$(bc)(\bar{b}\bar{n})$    &     &11911  &$\sim0\%$  &&$(bc)(\bar{b}\bar{s})$                &     &11963  &$\sim0\%$    \\
                        &$[7,5,3]$ &$(bc)(\bar{b}\bar{n})$    &     &12005  &$\sim0\%$  &&$(bc)(\bar{b}\bar{s})$                &     &12038  &$\sim0\%$    \\
                        &$[7,5,4]$ &$(bc)(\bar{b}\bar{n})$    &     &11929  &$\sim0\%$  &&$(bc)(\bar{b}\bar{s})$                &     &11981  &$\sim0\%$   \\
                        &$E_{CC1}$ &                                   &     &11327&             &&                                      &     &11429  &\\
                        &$E_{CC2}$ &                                   &     &11891&             &&                                      &     &11926  & \\
                        &$E_{CC}$ &                                    &     &11327&             &&                                      &     &11429  &\\
\midrule[1pt]
\multirow{9}*{$2^{+}$} &$[5,6,1]$ &$(\Upsilon D^{\ast})^{1}$         &11467&11469  &99.99\%   &&$(\Upsilon D_{s}^{\ast+})^{1}$         &11572&11573  &99.99\% \\
                       &$[6,6,1]$ &$(\Upsilon D^{\ast})^{8}$         &     &12071  &$\sim0\%$ &&$(\Upsilon D_{s}^{\ast+})^{8}$         &     &12123  &$\sim0\%$  \\
                       &$[5,6,2]$ &$(\bar{B^{\ast}}B_{c}^{\ast+})^{1}$ &11625&11627  &$\sim0\%$ &&$(\bar{B_{s}^{\ast}}B_{c}^{\ast+})^{1}$ &11715&11716  &$\sim0\%$  \\
                       &$[6,6,2]$ &$(\bar{B^{\ast}}B_{c}^{\ast+})^{8}$ &     &12012  &$\sim0\%$ &&$(\bar{B_{s}^{\ast}}B_{c}^{\ast+})^{8}$ &     &12048  &$\sim0\%$  \\
                       &$[7,6,3]$ &$(bc)(\bar{b}\bar{n})$    &     &12037  &$\sim0\%$ &&$(bc)(\bar{b}\bar{s})$                 &     &12078  &$\sim0\%$   \\
                       &$[7,6,4]$ &$(bc)(\bar{b}\bar{n})$    &     &11937  &$\sim0\%$ &&$(bc)(\bar{b}\bar{s})$                 &     &11992  &$\sim0\%$  \\
                       &$E_{CC1}$ &                                  &     &11469  &          &&                                       &     &11573  &    \\
                       &$E_{CC2}$ &                                  &     &11930  &          &&                                       &     &11982  &     \\
                       &$E_{CC}$ &                                   &     &11469  &          &&                                       &     &11573  &      \\
\bottomrule[1pt]
\end{tabular}
\end{center}
\end{table*}

\begin{table*}[ht]
\begin{center}
\caption{\label{part3} Contributions of each term in Hamiltonian to the energy of the $bc\bar{c}\bar{n}$ tetraquark and $bc\bar{b}\bar{n}$ tetraquark in ChQM.  $E_{M("channel")}$ stands for the sum of the theoretical thresholds of the lowest physical channel. (unit: MeV). }
  \renewcommand\arraystretch{1.5}
  \begin{tabular}{p{2.cm}<\centering p{1.5cm}<\centering p{1.5cm}<\centering p{1.5cm}<\centering p{1.5cm}<\centering <\centering p{0.5cm}<\centering p{1.2cm}<\centering p{1.5cm}<\centering p{1.5cm}<\centering p{1.5cm}<\centering}
 \toprule[1pt]
     & & $\langle H_{T}\rangle$ & $\langle V_{CON}\rangle$ & $\langle V_{OGE}\rangle $  & & & $\langle H_{T}\rangle $ & $\langle V_{CON}\rangle $ & $\langle V_{OGE} \rangle$ \\
    \cline{2-5} \cline{7-10}
\multirow{9}*{$J^{P}=1^{+}$}   &$E_{(D B_{c}^{\ast-})^{1}}$              &1662.7 &-1984.6 &-391.6 &&$E_{(\Upsilon D )^{1}}$ &1558.6 &-1432.2 &-1064.1 \\
   &$E_{CC1}$                                                           &1661.9 &-1985.3 &-391.0 &&$E_{CC1}$           &1558.4 &-1432.2 &-1063.9\\
   &$E_{CC2}$                                                           &1856.1 &-1870.1 &-312.9 &&$E_{CC2}$           &1772.2 &-1534.8 &-610.9\\
   &$E_{CC}$                                                            &1661.2 &-1998.9 &-384.0 &&$E_{CC}$            &1558.0 &-1432.1 &-1063.7 \\
   &$E_{M(D B_{c}^{\ast-})}$                                             &1660.2 &-1984.6 &-391.6 &&$E_{M(\Upsilon D)}$ &1556.6 &-1432.9 &-1064.1\\
   &$\Delta E_{(D B_{c}^{\ast-})^{1}}$                                   &2.5    &0.0     &0.0    &&$\Delta E_{(\Upsilon D)^{1}}$ &2.0  &0.7   &0.0 \\
   &$\Delta  E_{CC1}$                                                   &1.7    &-0.7    &0.6    &&$\Delta  E_{CC1}$                 &1.8  &0.7   &0.2\\
   &$\Delta  E_{CC2}$                                                   &195.9  &114.5   &78.7   &&$\Delta  E_{CC2}$                 &215.6&-101.9&453.2 \\
   &$\Delta  E_{CC}$                                                    &1.0    &-14.3   &7.6    &&$\Delta  E_{CC}$                  &1.4  &0.8   &-0.6\\
    \midrule[1pt]
  \multirow{9}*{$J^{P}=2^{+}$}  &$E_{(D^{\ast}B_{c}^{\ast-})^{1}}$             &1662.7 &-1984.6 &-249.6 &&$E_{(\Upsilon D^{\ast})^{1}}$   &1558.4 &-1432.2 &-922.1 \\
   &$E_{CC1}$                                                                 &1660.9 &-1984.9 &-249.1 &&$E_{CC1}$                   &1558.2 &-1432.1 &-921.9\\
   &$E_{CC2}$                                                                 &1852.5 &-1875.3 &-234.4 &&$E_{CC2}$                   &1724.6 &-1637.5 &-421.6\\
   &$E_{CC}$                                                                  &1660.3 &-1998.9 &-269.8 &&$E_{CC}$                    &1557.7 &-1431.8 &-921.3\\
   &$E_{M(D^{\ast}B_{c}^{\ast-})}$                                             &1660.2 &-1984.6 &-249.6 &&$E_{M(\Upsilon D^{\ast})}$  &1556.6 &-1432.9 &-922.8  \\
   &$\Delta E_{(D^{\ast}B_{c}^{\ast-})^{1}}$                                   &2.5    &0.0     &0.0    &&$\Delta E_{(\Upsilon D^{\ast})^{1}}$&1.8     &0.7   &0.7\\
   &$\Delta  E_{CC1}$                                                         &0.7    &-0.3    &0.5    &&$\Delta  E_{CC1}$                   &1.6     &0.8   &0.8\\
   &$\Delta  E_{CC2}$                                                         &192.3  &109.3   &15.2   &&$\Delta  E_{CC2}$                   &168.0   &-204.4&501.2\\
   &$\Delta  E_{CC}$                                                          &0.1    &-14.3   &-20.2  &&$\Delta  E_{CC}$                    &1.8     &0.7   &1.5 \\
  \bottomrule[1pt]
\end{tabular}
\end{center}
\end{table*}

\begin{table*}[ht]
\begin{center}
\caption{\label{part4} Contributions of each term in Hamiltonian to the energy of the $bc\bar{c}\bar{s}$  and $bc\bar{b}\bar{s}$ tetraquark in ChQM.  $E_{M("channel")}$ stands for the sum of the theoretical thresholds of the lowest physical channel. (unit: MeV). }
 \renewcommand\arraystretch{1.5}
  \begin{tabular}{p{2.cm}<\centering p{1.5cm}<\centering p{1.5cm}<\centering p{1.5cm}<\centering p{1.5cm}<\centering <\centering p{0.5cm}<\centering p{1.2cm}<\centering p{1.5cm}<\centering p{1.5cm}<\centering p{1.5cm}<\centering}
 \toprule[1pt]
     & & $\langle H_{T}\rangle$ & $\langle V_{CON}\rangle$ & $\langle V_{OGE}\rangle $  & & & $\langle H_{T}\rangle $ & $\langle V_{CON}\rangle $ & $\langle V_{OGE} \rangle$ \\
    \cline{2-5} \cline{7-10}
 \multirow{9}*{$J^{P}=(2^{+})$}    &$E_{(D_{s}^{\ast+}B_{c}^{\ast-})^{1}}$            &1176.7 &-1474.3 &-392.6 &&$E_{(\Upsilon D_{s}^{\ast+})^{1}}$ &1036.7 &-921.9  &-1065.1\\
   &$E_{CC1}$                                                                         &1176.3 &-1475.0 &-392.3 &&$E_{CC1}$                     &1036.5 &-921.9  &-1065.0\\
   &$E_{CC2}$                                                                         &1347.7 &-1396.2 &-347.2 &&$E_{CC2}$                     &1231.6 &-1100.5 &-636.5\\
   &$E_{CC}$                                                                          &1176.1 &-1478.2 &-391.9 &&$E_{CC}$                      &1036.5 &-921.9  &-1064.9\\
   &$E_{M(D_{s}^{\ast+}B_{c}^{\ast-})}$                                               &1174.9 &-1474.2 &-392.5 &&$E_{M(\Upsilon D_{s}^{\ast+})}$&1036.4 &-922.2  &-1065.6  \\
   &$\Delta E_{(D_{s}^{\ast+}B_{c}^{\ast-})^{1}}$                                     &1.8    &-0.1    &-0.1    &&$\Delta E_{(\Upsilon D_{s}^{\ast+})^{1}}$&0.3  &0.3   &0.5 \\
   &$\Delta  E_{CC1}$                                                                 &1.4    &-0.8    &0.2    &&$\Delta  E_{CC1}$                   &0.1  &0.3   &0.6\\
   &$\Delta  E_{CC2}$                                                                 &172.8  &78.0    &45.3    &&$\Delta  E_{CC2}$                   &195.2&-278.3&430.1\\
   &$\Delta  E_{CC}$                                                                  &1.2    &-4.0    &0.6    &&$\Delta  E_{CC}$                    &0.1  &0.3   &0.7\\
  \bottomrule[1pt]
\end{tabular}
\end{center}
\end{table*}

 Similar to the cases of $QQ\bar{Q}^\prime \bar{q}$ and $QQ^\prime \bar{Q} \bar{q}$ tetraquark states, we can estimate the eigenenergies of $QQ^\prime \bar{Q}\bar{q}$ tetraquark states. Our estimations of the eigenenergies of the $cb\bar{c}\bar{q^{}}$ and $bc\bar{b}\bar{q}$ tetraquark states are collected in Table \ref{cbcq} and \ref{bcbq}. From Table \ref{cbcq}, one can find the eigenenergies of $bc\bar{c}\bar{n}$ tetraquark state with $J^P=0^+$ obtained in the single channel estimations, the coupled channel estimations in each configuration and the full coupled channel estimations are all above the threshold of $D B_c^-$, which is 8140 MeV. Similarly, we also find that the eigenenergies of the $bc\bar{c}\bar{s}$ tetraquark states with $J^P=0^+$ are all above the threshold of $D_s^+ B_c^-$. As for $bc\bar{c} \bar{n}$ tetraquark state with $J^P=1^+$, we find that the eigenenergies obtained in the single channel estimations and the coupled channel estimations in the meson-meson and diquark-antidiquark configurations are all above the threshold of $D B_c^{\ast -}$, however, when considering the coupled channel effects in both meson-meson and diquark-antidiquark configurations, one obtains the eigenenergy to be 8159 MeV, which is 6 MeV below the threshold of  $D B_c^{\ast -}$. In this tetraquark state, the dominant component is $DB_c^{\ast -}$ with a percentage to be $91.57$. As for $bc\bar{c}\bar{s}$ tetraquark state with $J^P=1^+$, we find that the eigenenergies obtained in the single channel estimations, the coupled channel estimations in each configuration and the full coupled channel estimations are all above the threshold of $D_s^+ B_c^{\ast -}$.  As for the case of $J^p=2^+$, the eigenenergies of $bc\bar{c} \bar{n}$ and $bc\bar{c} \bar{s}$ obtained in the full coupled channel estimations are 8273 MeV and 8410 MeV, which are below the threshold of $D^\ast B_c^{\ast -}$ and $D_s^{\ast+} B_c^{\ast -}$, respectively. In the $bc\bar{b}\bar{n}$ tetraquark state with $J^P=2^+$, the dominant components are  $(D^\ast B_c^{\ast -})^1$,  $(J/\psi B^\ast)^8$ and $(bc)(\bar{c}\bar{n})$ with $[i,j,k]=[7,6,4]$, the corresponding percentages of these components are $72.10$, $11.04$ and $10.77$, respectively. As for $bc\bar{b}\bar{s}$ tetraquark state with $J^P=2^+$, the dominant component is $(D_s^{\ast +} B_c^{\ast-})^1$ with a percentage to be 95.29.

As for the $bc\bar{b}\bar{q}$ tetraquark system, the eigenenergies estimated in the ChQM  are collected in Table~\ref{bcbq}. From the table, one can find that the eigenenergies obtained in the single channel estimations, coupled channel estimations in each configuration and the full coupled channel estimations are all above the corresponding lowest physical threshold, which is different with the $bc\bar{c}\bar{q}$ tetraquark states, where one find three below threshold tetraquark states. To further analyze the role of the coupled channel effects, we estimate the average values of the operators in the Hamiltonian of $Q^\prime Q \bar{Q} \bar{q}$ system, which are collected in Tables \ref{part3} and \ref{part4}. From the tables, one can find the average values of kinetic terms increase when we include the interaction between mesons and coupled-channel effects. In the full coupled-channel estimations, we find the attraction from confinement potential becomes stronger for $bc\bar{c}\bar{n}$ tetraquark states with $J^P=1^+$ and $J^p=2^+$, but the attraction from the OGE potential becomes weak for the $bc\bar{c}\bar{n}$ tetraquark states with $J^P=1^+$, while this attraction becomes strong for the $bc\bar{c}\bar{n}$ tetraquark states with $J^P=2^+$. As for $bc\bar{c}\bar{s}$ tetraquark states, the full coupled-channel estimations indicate the average values of $H_T$, $V_{Con}$ and $V_{OGE}$ are close to those of $E_{M(\Upsilon D)}$, and the sum of these terms are positive. As for the $bc\bar{c}\bar{s}$ tetraquark state with $J^P=2^+$, the estimations indicate the confinement potential becomes strong in the full coupled-channel estimation.


\begin{table}[ht]
\begin{center}
\caption{\label{bound} Possible bound state with the different quantum number in ChQM. (unit: MeV). }
 \renewcommand\arraystretch{1.35}
  \begin{tabular}{p{1.cm}<\centering p{2.8cm}<\centering p{1.25cm}<\centering p{1.25cm}<\centering p{1.25cm}<\centering p{1.5cm}<\centering}
    \toprule[1pt]
    $J^{P}$& quark components    & $E_{th}$ & $E_{cc}$ & $B_{cc}$ \\
   \midrule[1pt]
    $2^{+}$& $cc\bar{c}\bar{n}$ &5104       & 5095    &-9 \\
    $0^{+}$& $bb\bar{c}\bar{n}$ &11554      &11552    &-2 \\
    $1^{+}$& $bb\bar{c}\bar{n}$ &11579      &11566    &-13 \\
    $2^{+}$& $bb\bar{c}\bar{n}$ &11625      &11613    &-12 \\
    $1^{+}$& $bc\bar{c}\bar{n}$ &8165       &8159     &-6 \\
    $2^{+}$& $bc\bar{c}\bar{n}$ &8307       &8273     &-34 \\
    $2^{+}$& $bc\bar{c}\bar{s}$ &8412       &8410     &-2 \\

  \bottomrule[1pt]
\end{tabular}
\end{center}
\end{table}

\section{Summary}\label{Sum}

To summarize, inspired by the recent observation of fully heavy tetraquark states, we perform a systematic estimation of the triply tetraquark states in a chiral quark model,  where the coupled channel effects of meson-meson configuration and diquark-antidiquark configurations are included. The eigenenergies of the $S$-wave ground states have been estimated. After including the coupled channel effects of both configurations, We notice that the eigenenergies of some tetraquark states are below the corresponding lowest threshold of the physical channel, which indicates that these tetraquark states cannot fall apart directly and thus are stable for strong decay. In Table~\ref{bound}, we collect all the stable tetraquark states estimated in the present work. For comparison, we also list the corresponding lowest thresholds of the physical channel.

Moreover, comparing with the results in Refs.~\cite{Lu:2021kut,Weng:2021ngd,Jiang:2017tdc,Chen:2016ont}, we find that the masses of the diquark-antidiquark configurations are several hundred MeV higher than those of the color-magnetic interaction model~\cite{Weng:2021ngd,Chen:2016ont} and QCD sum rules~\cite{Jiang:2017tdc}, while the masses under an extended relativized quark model~\cite{Lu:2021kut} are generally consistent with present estimations of the diquark-diquark configurations. Although there are discrepancies in the estimated masses due to different input parameters and different interactions in different models, the conclusions are basically the same for the triply heavy tetraquark system, i.e., no stable states are found in the diquark-antidiquark configurations except for the estimation of QCD sum rules~\cite{Jiang:2017tdc}. But when we consider the coupled-channel effects of diquark-antidiquark and meson-meson configurations simultaneously, we find there exist several stable tetraquark states which are below the corresponding lowest physical threshold, which may accessible for experiments in LHCb.

\acknowledgments{This work is supported partly by the National Natural Science Foundation of China under
Contract Nos. 12175037, 11775050, 11775118, and 11535005, and this work is also supported by China Postdoctoral Science Foundation funded project No. 2021M690626, No. 1107020201, and the Fundamental Research Funds for the Central Universities No.2242022R20040.}

\appendix

\section{The wave function of the triply heavy tetraquark}
\label{Sec:Appendix}

\subsubsection{The color wave function}

For the meson-meson configurations, the color wave functions of a $q\bar{q}$ cluster are,
\begin{eqnarray}
\nonumber
C^{1}_{[111]} &=& \sqrt{\frac{1}{3}}(r\bar{r}+g\bar{g}+b\bar{b}), \\ \nonumber
C^{2}_{[21]} &=& r\bar{b},\hspace{1cm} C^{3}_{[21]} =  -r\bar{g},                    \\ \nonumber
C^{4}_{[21]} &=& g\bar{b}, \hspace{1cm}C^{5}_{[21]} =  -b\bar{g},              \\ \nonumber
C^{6}_{[21]} &=& g\bar{r}, \hspace{1cm}C^{7}_{[21]} =   b\bar{r},         \\ \nonumber
C^{8}_{[21]} &=&  \sqrt{\frac{1}{2}}(r\bar{r}-g\bar{g}),       \\ \nonumber
C^{9}_{[21]} &=&  \sqrt{\frac{1}{6}}(-r\bar{r}-g\bar{g}+2b\bar{b}), \\
\end{eqnarray}
where the subscript [111] and [21] stand for color-singlet ($\textbf{1}_{c}$) and color-octet ($\textbf{8}_{c}$), respectively. Then the color-singlet tetraquark $\mathrm{SU(3)_{color}}$ wave functions can be constructed by two color-singlet clusters, i.e.,$\textbf{1}_{c}\otimes\textbf{1}_{c}$) and by two color-octet clusters, i.e., $\textbf{8}_{c}\otimes\textbf{8}_{c}$), which are,
\begin{equation}
\chi^{c}_{1} = C^{1}_{[111]}C^{1}_{[111]},\nonumber\\
\end{equation}
\begin{equation}
\begin{split}
  \chi^{c}_{2} =&\sqrt{\frac{1}{8}}(C^{2}_{[21]}C^{7}_{[21]}-C^{4}_{[21]}C^{5}_{[21]}-C^{3}_{[21]}C^{6}_{[21]}\\
                &+C^{8}_{[21]}C^{8}_{[21]}-C^{6}_{[21]}C^{3}_{[21]}+C^{9}_{[21]}C^{9}_{[21]}\\
                &-C^{5}_{[21]}C^{4}_{[21]}+C^{7}_{[21]}C^{2}_{[21]}).
\end{split}
\end{equation}

For the diquark-antidiquark configuration, the color wave functions of the diquark clusters are,
\begin{eqnarray}
\nonumber
  C^{1}_{[2]} &=& rr,  C^{2}_{[2]} = \sqrt{\frac{1}{2}}(rg+gr), \\ \nonumber
  C^{3}_{[2]} &=& gg,  C^{4}_{[2]} = \sqrt{\frac{1}{2}}(rb+br),\\ \nonumber
  C^{5}_{[2]} &=& \sqrt{\frac{1}{2}}(gb+bg),  C^{6}_{[2]} = bb, \\ \nonumber
  C^{7}_{[11]} &=& \sqrt{\frac{1}{2}}(rg-gr),  C^{8}_{[11]} = \sqrt{\frac{1}{2}}(rb-br), \\ \nonumber
  C^{9}_{[11]} &=&\sqrt{\frac{1}{2}}(gb-bg). \\
\end{eqnarray}
While the color wave functions of the antidiquark clusters are,
\begin{eqnarray}
\nonumber
  C^{1}_{[22]} &=& \bar{r}\bar{r},  C^{2}_{[22]} = -\sqrt{\frac{1}{2}}(\bar{r}\bar{g}+\bar{g}\bar{r}),\\  \nonumber
  C^{3}_{[22]} &=& \bar{g}\bar{g},  C^{4}_{[22]} = \sqrt{\frac{1}{2}}(\bar{r}\bar{b}+\bar{b}\bar{r}), \\  \nonumber
  C^{5}_{[22]} &=& -\sqrt{\frac{1}{2}}(\bar{g}\bar{b}+\bar{b}\bar{g}), C^{6}_{[22]} = \bar{b}\bar{b}, \\ \nonumber
  C^{7}_{[211]}&=& \sqrt{\frac{1}{2}}(\bar{r}\bar{g}-\bar{g}\bar{r}), C^{8}_{[211]} = -\sqrt{\frac{1}{2}}(\bar{r}\bar{b}-\bar{b}\bar{r}), \\ \nonumber
  C^{9}_{[211]} &=& \sqrt{\frac{1}{2}}(\bar{g}\bar{b}-\bar{b}\bar{g}). \\
\end{eqnarray}
The color-singlet wave functions of the diquark-antidiquark configuration can be the product of color sextet and antisextet clusters ($\textbf{6}_{c}\otimes\bar{\textbf{6}}_{c}$) or the product of color-triplet and antitriplet cluster ($\textbf{6}_{c}\otimes\bar{\textbf{6}}_{c}$), which read,
\begin{equation}
\begin{split}
\chi^{c}_{3} = &\sqrt{\frac{1}{6}}(C^{1}_{[2]}C^{1}_{[22]}-C^{2}_{[2]}C^{[2]}_{[22]}+C^{3}_{[2]}C^{3}_{[22]} \\
               &+C^{4}_{[2]}C^{4}_{[22]}-C^{5}_{[2]}C^{5}_{[22]}+C^{6}_{2}C^{6}_{22}),\nonumber
\end{split}
\end{equation}

\begin{equation}
\begin{split}
  \chi^{c}_{4} =&\sqrt{\frac{1}{3}}(C^{7}_{[11]}C^{7}_{[211]}-C^{8}_{[11]}C^{8}_{[211]}+C^{9}_{[11]}C^{9}_{[211]}).
\end{split}
\end{equation}


\subsubsection{The flavor wave function}
For the flavor degree of freedom, the quark content of the investigated 4-quark system is $QQ\bar{Q}\bar{q},\  Q=\{c, b\},\ q=\{u, d, s\}$, the isospin could be $1/2$ and $0$. Here, we adopt $F^{i}_m$ and $F^{i}_d$ to denote the flavor wave functions of the tetraquark system in the meson-meson and diquark-antidiquark configurations, respectively. In the present work, the flavor wave function of the $QQ\bar{Q}\bar{q}$ system can be categorized into three types, which are $QQ\bar{Q}\bar{q}, QQ\bar{Q^{\prime}}\bar{q}$ and $QQ^{\prime}\bar{Q}\bar{q}$, respectively.

 For the $QQ\bar{Q}\bar{q}$ system, the flavor wave functions can be,
\begin{eqnarray}
  F^{1}_{m}&=& (Q\bar{Q})(Q\bar{q}),   \qquad F^{2}_{d}= (QQ)(\bar{Q}\bar{q}),
\end{eqnarray}
and for the $QQ\bar{Q^{\prime}}\bar{q}$ system, the flavor wave functions can be read as,
\begin{eqnarray}
  F^{3}_{m}&=& (Q\bar{Q^{\prime}})(Q\bar{q}),   \qquad F^{4}_{d}= (QQ)(\bar{Q^{\prime}}\bar{q}) .
\end{eqnarray}
While the flavor wave functions for the $QQ^{\prime}\bar{Q}\bar{q}$ system read,
\begin{eqnarray}
   F^{5}_{m}&=& (Q\bar{Q})(Q^{\prime}\bar{q}),\qquad F^{6}_{m}= (Q\bar{q})(Q^{\prime}\bar{Q}),\nonumber\\
    \qquad F^{7}_{d} &=& (QQ^{\prime})(\bar{Q}\bar{q}) .
   \end{eqnarray}

\subsubsection{The spin wave function}
The total spin $S$ of tetraquark states can be 0, 1, and 2.
The spin wave functions of two body clusters are
\begin{eqnarray}
\nonumber \chi_{11}&=& \alpha\alpha,\\
\nonumber \chi_{10} &=& \sqrt{\frac{1}{2}}(\alpha\beta+\beta\alpha),\\
\nonumber \chi_{1-1} &=& \beta\beta, \\
            \chi_{00} &=& \sqrt{\frac{1}{2}}(\alpha\beta-\beta\alpha).
\end{eqnarray}
Then, the spin wave functions of the tetraquark state $S^{i}_{s}$ can be constructed by considering the coupling of two subcluster spin wave functions with SU(2) algebra, which read,
\begin{eqnarray}
\nonumber S^{1}_{0}&=&\chi_{00}\chi_{00},\\
\nonumber S^{2}_{0}&=&\sqrt{\frac{1}{3}}(\chi_{11}\chi_{1-1}-\chi_{10}\chi_{10}+\chi_{1-1}\chi_{11}),\\
\nonumber S^{3}_{1}&=&\chi_{00}\chi_{11},\\
\nonumber S^{4}_{1}&=&\chi_{11}\chi_{00},\\
\nonumber S^{5}_{1}&=&\sqrt{\frac{1}{2}}(\chi_{11}\chi_{10}-\chi_{10}\chi_{11}),\\
S^{6}_{2}&=&\chi_{11}\chi_{11}.
\end{eqnarray}


\begin{thebibliography}{99}
\bibitem{Chen:2016qju}
H.~X.~Chen, W.~Chen, X.~Liu and S.~L.~Zhu,
Phys. Rept. \textbf{639} (2016), 1-121
doi:10.1016/j.physrep.2016.05.004
[arXiv:1601.02092 [hep-ph]].

\bibitem{Swanson:2006st}
E.~S.~Swanson,
Phys. Rept. \textbf{429} (2006), 243-305
doi:10.1016/j.physrep.2006.04.003
[arXiv:hep-ph/0601110 [hep-ph]].

\bibitem{Voloshin:2007dx}
M.~B.~Voloshin,
Prog. Part. Nucl. Phys. \textbf{61} (2008), 455-511
doi:10.1016/j.ppnp.2008.02.001
[arXiv:0711.4556 [hep-ph]].

\bibitem{Chen:2016heh}
R.~Chen, X.~Liu and S.~L.~Zhu,
Nucl. Phys. A \textbf{954} (2016), 406-421
doi:10.1016/j.nuclphysa.2016.04.012
[arXiv:1601.03233 [hep-ph]].

\bibitem{Esposito:2016noz}
A.~Esposito, A.~Pilloni and A.~D.~Polosa,
Phys. Rept. \textbf{668} (2017), 1-97
doi:10.1016/j.physrep.2016.11.002
[arXiv:1611.07920 [hep-ph]].

\bibitem{Lebed:2016hpi}
R.~F.~Lebed, R.~E.~Mitchell and E.~S.~Swanson,
Prog. Part. Nucl. Phys. \textbf{93} (2017), 143-194
doi:10.1016/j.ppnp.2016.11.003
[arXiv:1610.04528 [hep-ph]].

\bibitem{Guo:2017jvc}
F.~K.~Guo, C.~Hanhart, U.~G.~Mei\ss{}ner, Q.~Wang, Q.~Zhao and B.~S.~Zou,
Rev. Mod. Phys. \textbf{90} (2018) no.1, 015004
doi:10.1103/RevModPhys.90.015004
[arXiv:1705.00141 [hep-ph]].


\bibitem{Gell-Mann:1964ewy}
M.~Gell-Mann,
Phys. Lett. \textbf{8} (1964), 214-215
doi:10.1016/S0031-9163(64)92001-3

\bibitem{Belle:2003nnu}
S.~K.~Choi \textit{et al.} [Belle],
Phys. Rev. Lett. \textbf{91} (2003), 262001
doi:10.1103/PhysRevLett.91.262001
[arXiv:hep-ex/0309032 [hep-ex]].

\bibitem{BaBar:2003oey}
B.~Aubert \textit{et al.} [BaBar],
Phys. Rev. Lett. \textbf{90} (2003), 242001
doi:10.1103/PhysRevLett.90.242001
[arXiv:hep-ex/0304021 [hep-ex]].


\bibitem{CLEO:2003ggt}
D.~Besson \textit{et al.} [CLEO],
Phys. Rev. D \textbf{68} (2003), 032002
[erratum: Phys. Rev. D \textbf{75} (2007), 119908]
doi:10.1103/PhysRevD.68.032002
[arXiv:hep-ex/0305100 [hep-ex]].

\bibitem{Belle:2003guh}
P.~Krokovny \textit{et al.} [Belle],
Phys. Rev. Lett. \textbf{91}, 262002 (2003)
doi:10.1103/PhysRevLett.91.262002
[arXiv:hep-ex/0308019 [hep-ex]].

\bibitem{Godfrey:2003kg}
S.~Godfrey,
Phys. Lett. B \textbf{568}, 254-260 (2003)
doi:10.1016/j.physletb.2003.06.049
[arXiv:hep-ph/0305122 [hep-ph]].


\bibitem{Rosner:2006jz}
J.~L.~Rosner,
J. Phys. G \textbf{34}, S127-S148 (2007)
doi:10.1088/0954-3899/34/7/S07
[arXiv:hep-ph/0609195 [hep-ph]].


\bibitem{Godfrey:1985xj}
S.~Godfrey and N.~Isgur,
Phys. Rev. D \textbf{32}, 189-231 (1985)
doi:10.1103/PhysRevD.32.189


\bibitem{Liu:2020ruo}
J.~Liu, Q.~Wu, J.~He, D.~Y.~Chen and T.~Matsuki,
Phys. Rev. D \textbf{101} (2020) no.1, 014003
doi:10.1103/PhysRevD.101.014003
[arXiv:2001.00212 [hep-ph]].


\bibitem{Xiao:2016hoa}
c.~J.~Xiao, D.~Y.~Chen and Y.~L.~Ma,
Phys. Rev. D \textbf{93} (2016) no.9, 094011
doi:10.1103/PhysRevD.93.094011
[arXiv:1601.06399 [hep-ph]].


\bibitem{Cheng:2003kg}
H.~Y.~Cheng and W.~S.~Hou,
Phys. Lett. B \textbf{566}, 193-200 (2003)
doi:10.1016/S0370-2693(03)00834-7
[arXiv:hep-ph/0305038 [hep-ph]].

\bibitem{Chen:2004dy}
Y.~Q.~Chen and X.~Q.~Li,
Phys. Rev. Lett. \textbf{93}, 232001 (2004).
doi:10.1103/PhysRevLett.93.232001
[arXiv:hep-ph/0407062 [hep-ph]].

\bibitem{Kim:2005gt}
H.~Kim and Y.~Oh,
Phys. Rev. D \textbf{72}, 074012 (2005).
doi:10.1103/PhysRevD.72.074012
[arXiv:hep-ph/0508251 [hep-ph]].

\bibitem{Nielsen:2005ia}
M.~Nielsen, R.~D.~Matheus, F.~S.~Navarra, M.~E.~Bracco and A.~Lozea,
Nucl. Phys. B Proc. Suppl. \textbf{161}, 193-199 (2006).
doi:10.1016/j.nuclphysbps.2006.08.045
[arXiv:hep-ph/0509131 [hep-ph]].

\bibitem{Terasaki:2005kc}
K.~Terasaki,
[arXiv:hep-ph/0512285 [hep-ph]].

\bibitem{Wang:2006uba}
Z.~G.~Wang and S.~L.~Wan,
Nucl. Phys. A \textbf{778}, 22-29 (2006).
doi:10.1016/j.nuclphysa.2006.07.041
[arXiv:hep-ph/0602080 [hep-ph]].


\bibitem{Barnes:2003dj}
T.~Barnes, F.~E.~Close and H.~J.~Lipkin,
Phys. Rev. D \textbf{68}, 054006 (2003).
doi:10.1103/PhysRevD.68.054006
[arXiv:hep-ph/0305025 [hep-ph]].

\bibitem{Navarra:2015iea}
F.~S.~Navarra, M.~Nielsen, E.~Oset and T.~Sekihara,
Phys. Rev. D \textbf{92}, no.1, 014031 (2015).
doi:10.1103/PhysRevD.92.014031
[arXiv:1501.03422 [hep-ph]].

\bibitem{Kolomeitsev:2003ac}
E.~E.~Kolomeitsev and M.~F.~M.~Lutz,
Phys. Lett. B \textbf{582}, 39-48 (2004).
doi:10.1016/j.physletb.2003.10.118
[arXiv:hep-ph/0307133 [hep-ph]].

\bibitem{Hofmann:2003je}
J.~Hofmann and M.~F.~M.~Lutz,
Nucl. Phys. A \textbf{733}, 142-152 (2004).
doi:10.1016/j.nuclphysa.2003.12.013
[arXiv:hep-ph/0308263 [hep-ph]].

\bibitem{Guo:2006fu}
F.~K.~Guo, P.~N.~Shen, H.~C.~Chiang, R.~G.~Ping and B.~S.~Zou,
Phys. Lett. B \textbf{641}, 278-285 (2006).
doi:10.1016/j.physletb.2006.08.064
[arXiv:hep-ph/0603072 [hep-ph]].

\bibitem{Zhang:2006ix}
Y.~J.~Zhang, H.~C.~Chiang, P.~N.~Shen and B.~S.~Zou,
Phys. Rev. D \textbf{74}, 014013 (2006).
doi:10.1103/PhysRevD.74.014013
[arXiv:hep-ph/0604271 [hep-ph]].

\bibitem{Rosner:2006vc}
J.~L.~Rosner,
Phys. Rev. D \textbf{74}, 076006 (2006).
doi:10.1103/PhysRevD.74.076006
[arXiv:hep-ph/0608102 [hep-ph]].

\bibitem{Guo:2006rp}
F.~K.~Guo, P.~N.~Shen and H.~C.~Chiang,
Phys. Lett. B \textbf{647}, 133-139 (2007).
doi:10.1016/j.physletb.2007.01.050
[arXiv:hep-ph/0610008 [hep-ph]].




\bibitem{Aubert:2004fc}
  B.~Aubert {\it et al.} [BaBar Collaboration],
  Phys.\ Rev.\ Lett.\  {\bf 93}, 041801 (2004)
  doi:10.1103/PhysRevLett.93.041801
  [hep-ex/0402025].

\bibitem{Aubert:2004ns}
  B.~Aubert {\it et al.} [BaBar Collaboration],
  Phys.\ Rev.\ D {\bf 71}, 071103 (2005)
  doi:10.1103/PhysRevD.71.071103
  [hep-ex/0406022].


\bibitem{Aubert:2005eg}
  B.~Aubert {\it et al.} [BaBar Collaboration],
  Phys.\ Rev.\ D {\bf 71}, 052001 (2005)
  doi:10.1103/PhysRevD.71.052001
  [hep-ex/0502025].

\bibitem{Aubert:2005zh}
  B.~Aubert {\it et al.} [BaBar Collaboration],
  Phys.\ Rev.\ D {\bf 73}, 011101 (2006)
  doi:10.1103/PhysRevD.73.011101
  [hep-ex/0507090].

\bibitem{Aubert:2005vi}
  B.~Aubert {\it et al.} [BaBar Collaboration],
  Phys.\ Rev.\ Lett.\  {\bf 96}, 052002 (2006)
  doi:10.1103/PhysRevLett.96.052002
  [hep-ex/0510070].

\bibitem{Aubert:2006aj}
  B.~Aubert {\it et al.} [BaBar Collaboration],
  Phys.\ Rev.\ D {\bf 74}, 071101 (2006)
  doi:10.1103/PhysRevD.74.071101
  [hep-ex/0607050].

\bibitem{Aubert:2007rva}
  B.~Aubert {\it et al.} [BaBar Collaboration],
  Phys.\ Rev.\ D {\bf 77}, 011102 (2008)
  doi:10.1103/PhysRevD.77.011102
  [arXiv:0708.1565 [hep-ex]].

\bibitem{Aubert:2008gu}
  B.~Aubert {\it et al.} [BaBar Collaboration],
  Phys.\ Rev.\ D {\bf 77}, 111101 (2008)
  doi:10.1103/PhysRevD.77.111101
  [arXiv:0803.2838 [hep-ex]].

\bibitem{Aubert:2008ae}
  B.~Aubert {\it et al.} [BaBar Collaboration],
  Phys.\ Rev.\ Lett.\  {\bf 102}, 132001 (2009)
  doi:10.1103/PhysRevLett.102.132001
  [arXiv:0809.0042 [hep-ex]].

\bibitem{delAmoSanchez:2010jr}
  P.~del Amo Sanchez {\it et al.} [BaBar Collaboration],
  Phys.\ Rev.\ D {\bf 82}, 011101 (2010)
  doi:10.1103/PhysRevD.82.011101
  [arXiv:1005.5190 [hep-ex]].

\bibitem{Acosta:2003zx}
  D.~Acosta {\it et al.} [CDF Collaboration],
  Phys.\ Rev.\ Lett.\  {\bf 93}, 072001 (2004)
  doi:10.1103/PhysRevLett.93.072001
  [hep-ex/0312021].

\bibitem{Abulencia:2005zc}
  A.~Abulencia {\it et al.} [CDF Collaboration],
  Phys.\ Rev.\ Lett.\  {\bf 96}, 102002 (2006)
  doi:10.1103/PhysRevLett.96.102002
  [hep-ex/0512074].

\bibitem{Abulencia:2006ma}
  A.~Abulencia {\it et al.} [CDF Collaboration],
  Phys.\ Rev.\ Lett.\  {\bf 98}, 132002 (2007)
  doi:10.1103/PhysRevLett.98.132002
  [hep-ex/0612053].

\bibitem{Aaltonen:2009vj}
  T.~Aaltonen {\it et al.} [CDF Collaboration],
  Phys.\ Rev.\ Lett.\  {\bf 103}, 152001 (2009)
  doi:10.1103/PhysRevLett.103.152001
  [arXiv:0906.5218 [hep-ex]].

\bibitem{Abazov:2004kp}
  V.~M.~Abazov {\it et al.} [D0 Collaboration],
  Phys.\ Rev.\ Lett.\  {\bf 93}, 162002 (2004)
  doi:10.1103/PhysRevLett.93.162002
  [hep-ex/0405004].

\bibitem{CMS:2011yra}
  [CMS Collaboration],
  CMS-PAS-BPH-10-018.

\bibitem{Vesentini:2012lea}
  A.~Vesentini [CMS Collaboration],
  Nuovo Cim.\ C {\bf 035}, no. 05, 21 (2012).
  doi:10.1393/ncc/i2012-11306-6

\bibitem{Chatrchyan:2013cld}
  S.~Chatrchyan {\it et al.} [CMS Collaboration],
  JHEP {\bf 1304}, 154 (2013)
  doi:10.1007/JHEP04(2013)154
  [arXiv:1302.3968 [hep-ex]].

\bibitem{DallOsso:2013rtt}
  M.~Dall'Osso [CMS Collaboration],
  PoS Beauty {\bf 2013}, 066 (2013).
  doi:10.22323/1.190.0066


\bibitem{DallOsso:2014cmg}
  M.~Dall'Osso [CMS Collaboration],
  Nuovo Cim.\ C {\bf 037}, no. 01, 283 (2014).
  doi:10.1393/ncc/i2014-11709-3

\bibitem{Sirunyan:2020qir}
  A.~M.~Sirunyan {\it et al.} [CMS Collaboration],
  Phys.\ Rev.\ Lett.\  {\bf 125}, no. 15, 152001 (2020)
  doi:10.1103/PhysRevLett.125.152001
  [arXiv:2005.04764 [hep-ex]].

\bibitem{Aaij:2011sn}
  R.~Aaij {\it et al.} [LHCb Collaboration],
  Eur.\ Phys.\ J.\ C {\bf 72}, 1972 (2012)
  doi:10.1140/epjc/s10052-012-1972-7
  [arXiv:1112.5310 [hep-ex]].


\bibitem{LHCb:2011bia}
  [LHCb Collaboration],
  LHCb-CONF-2011-043, CERN-LHCb-CONF-2011-043.

\bibitem{LHCb:2011cra}
  [LHCb Collaboration],
  LHCb-CONF-2011-021, CERN-LHCb-CONF-2011-021.

\bibitem{Aaij:2013zoa}
  R.~Aaij {\it et al.} [LHCb Collaboration],
  Phys.\ Rev.\ Lett.\  {\bf 110}, 222001 (2013)
  doi:10.1103/PhysRevLett.110.222001
  [arXiv:1302.6269 [hep-ex]].

\bibitem{Aaij:2013rha}
  R.~Aaij {\it et al.} [LHCb Collaboration],
  Eur.\ Phys.\ J.\ C {\bf 73}, no. 6, 2462 (2013)
  doi:10.1140/epjc/s10052-013-2462-2
  [arXiv:1303.7133 [hep-ex]].

\bibitem{Aaij:2014ala}
  R.~Aaij {\it et al.} [LHCb Collaboration],
  Nucl.\ Phys.\ B {\bf 886}, 665 (2014)
  doi:10.1016/j.nuclphysb.2014.06.011
  [arXiv:1404.0275 [hep-ex]].

\bibitem{Aaij:2015eva}
  R.~Aaij {\it et al.} [LHCb Collaboration],
  Phys.\ Rev.\ D {\bf 92}, no. 1, 011102 (2015)
  doi:10.1103/PhysRevD.92.011102
  [arXiv:1504.06339 [hep-ex]].

\bibitem{Aaij:2016kxn}
  R.~Aaij {\it et al.} [LHCb Collaboration],
  Phys.\ Lett.\ B {\bf 769}, 305 (2017)
  doi:10.1016/j.physletb.2017.03.046
  [arXiv:1607.06446 [hep-ex]].

\bibitem{Aaij:2017tzn}
  R.~Aaij {\it et al.} [LHCb Collaboration],
  Eur.\ Phys.\ J.\ C {\bf 77}, no. 9, 609 (2017)
  doi:10.1140/epjc/s10052-017-5151-8
  [arXiv:1706.07013 [hep-ex]].

\bibitem{Aaij:2019zkm}
  R.~Aaij {\it et al.} [LHCb Collaboration],
  JHEP {\bf 1909}, 028 (2019)
  doi:10.1007/JHEP09(2019)028
  [arXiv:1907.00954 [hep-ex]].


\bibitem{Durham:2020zuw}
  J.~Matthew Durham [LHCb Collaboration],
  arXiv:2002.01551 [hep-ex].

\bibitem{Aaij:2020qga}
  R.~Aaij {\it et al.} [LHCb Collaboration],
  Phys.\ Rev.\ D {\bf 102}, no. 9, 092005 (2020)
  doi:10.1103/PhysRevD.102.092005
  [arXiv:2005.13419 [hep-ex]].

\bibitem{Aaij:2020xjx}
  R.~Aaij {\it et al.} [LHCb Collaboration],
  JHEP {\bf 2008}, 123 (2020)
  doi:10.1007/JHEP08(2020)123
  [arXiv:2005.13422 [hep-ex]].

\bibitem{Aaij:2020tzn}
  R.~Aaij {\it et al.} [LHCb Collaboration],
  arXiv:2011.01867 [hep-ex].

\bibitem{Ablikim:2013dyn}
  M.~Ablikim {\it et al.} [BESIII Collaboration],
  Phys.\ Rev.\ Lett.\  {\bf 112}, no. 9, 092001 (2014)
  doi:10.1103/PhysRevLett.112.092001
  [arXiv:1310.4101 [hep-ex]].

\bibitem{Ablikim:2019soz}
  M.~Ablikim {\it et al.} [BESIII Collaboration],
  Phys.\ Rev.\ Lett.\  {\bf 122}, no. 20, 202001 (2019)
  doi:10.1103/PhysRevLett.122.202001
  [arXiv:1901.03992 [hep-ex]].

\bibitem{Ablikim:2019zio}
  M.~Ablikim {\it et al.} [BESIII Collaboration],
  Phys.\ Rev.\ Lett.\  {\bf 122}, no. 23, 232002 (2019)
  doi:10.1103/PhysRevLett.122.232002
  [arXiv:1903.04695 [hep-ex]].

\bibitem{Ablikim:2020xpq}
  M.~Ablikim {\it et al.} [BESIII Collaboration],
  Phys.\ Rev.\ Lett.\  {\bf 124}, no. 24, 242001 (2020)
  doi:10.1103/PhysRevLett.124.242001
  [arXiv:2001.01156 [hep-ex]].



\bibitem{Barnes:2003vb}
  T.~Barnes and S.~Godfrey,
  Phys.\ Rev.\ D {\bf 69}, 054008 (2004)
  doi:10.1103/PhysRevD.69.054008
  [hep-ph/0311162].

\bibitem{Eichten:2004uh}
  E.~J.~Eichten, K.~Lane and C.~Quigg,
  Phys.\ Rev.\ D {\bf 69}, 094019 (2004)
  doi:10.1103/PhysRevD.69.094019
  [hep-ph/0401210].

\bibitem{Chen:2007vu}
  Y.~Chen {\it et al.} [CLQCD Collaboration],
  hep-lat/0701021 [HEP-LAT].

\bibitem{Liu:2007uj}
  X.~Liu and Y.~M.~Wang,
  Eur.\ Phys.\ J.\ C {\bf 49}, 643 (2007).
  doi:10.1140/epjc/s10052-006-0135-0

\bibitem{Wang:2010ej}
  T.~H.~Wang and G.~L.~Wang,
  Phys.\ Lett.\ B {\bf 697}, 233 (2011)
  doi:10.1016/j.physletb.2011.02.014
  [arXiv:1006.3363 [hep-ph]].

\bibitem{Kalashnikova:2010hv}
  Y.~S.~Kalashnikova and A.~V.~Nefediev,
  Phys.\ Rev.\ D {\bf 82}, 097502 (2010)
  doi:10.1103/PhysRevD.82.097502
  [arXiv:1008.2895 [hep-ph]].

\bibitem{Wang:2012cp}
  T.~Wang, G.~L.~Wang, Y.~Jiang and W.~L.~Ju,
  J.\ Phys.\ G {\bf 40}, 035003 (2013)
  doi:10.1088/0954-3899/40/3/035003
  [arXiv:1205.5725 [hep-ph]].


\bibitem{Meng:2007cx}
C.~Meng and K.~T.~Chao,
Phys. Rev. D \textbf{75} (2007), 114002
doi:10.1103/PhysRevD.75.114002
[arXiv:hep-ph/0703205 [hep-ph]].


\bibitem{Liu:2008qb}
  Y.~R.~Liu and Z.~Y.~Zhang,
  Phys.\ Rev.\ C {\bf 79}, 035206 (2009)
  doi:10.1103/PhysRevC.79.035206
  [arXiv:0805.1616 [hep-ph]].

\bibitem{Gamermann:2009fv}
  D.~Gamermann and E.~Oset,
  Phys.\ Rev.\ D {\bf 80}, 014003 (2009)
  doi:10.1103/PhysRevD.80.014003
  [arXiv:0905.0402 [hep-ph]].


\bibitem{Voloshin:1976ap}
  M.~B.~Voloshin and L.~B.~Okun,
  JETP Lett.\  {\bf 23}, 333 (1976)
  [Pisma Zh.\ Eksp.\ Teor.\ Fiz.\  {\bf 23}, 369 (1976)].

\bibitem{DeRujula:1976zlg}
  A.~De Rujula, H.~Georgi and S.~L.~Glashow,
  Phys.\ Rev.\ Lett.\  {\bf 38}, 317 (1977).
  doi:10.1103/PhysRevLett.38.317

\bibitem{Tornqvist:1993ng}
  N.~A.~Tornqvist,
  Z.\ Phys.\ C {\bf 61}, 525 (1994)
  doi:10.1007/BF01413192
  [hep-ph/9310247].




\bibitem{Thomas:2008ja}
  C.~E.~Thomas and F.~E.~Close,
  Phys.\ Rev.\ D {\bf 78}, 034007 (2008)
  doi:10.1103/PhysRevD.78.034007
  [arXiv:0805.3653 [hep-ph]].

\bibitem{Lee:2009hy}
  I.~W.~Lee, A.~Faessler, T.~Gutsche and V.~E.~Lyubovitskij,
  Phys.\ Rev.\ D {\bf 80}, 094005 (2009)
  doi:10.1103/PhysRevD.80.094005
  [arXiv:0910.1009 [hep-ph]].

\bibitem{Chen:2009zzi}
  X.~Chen, B.~Wang, X.~Li, X.~Zeng, S.~Yu and X.~Lu,
  Phys.\ Rev.\ D {\bf 79}, 114006 (2009).
  doi:10.1103/PhysRevD.79.114006

\bibitem{Gamermann:2009uq}
  D.~Gamermann, J.~Nieves, E.~Oset and E.~Ruiz Arriola,
  Phys.\ Rev.\ D {\bf 81}, 014029 (2010)
  doi:10.1103/PhysRevD.81.014029
  [arXiv:0911.4407 [hep-ph]].

\bibitem{Ortega:2010qq}
  P.~G.~Ortega, J.~Segovia, D.~R.~Entem and F.~Fernandez,
  AIP Conf.\ Proc.\  {\bf 1257}, no. 1, 331 (2010)
  doi:10.1063/1.3483344
  [arXiv:1001.3948 [hep-ph]].

\bibitem{Guo:2013sya}
  F.~K.~Guo, C.~Hidalgo-Duque, J.~Nieves and M.~P.~Valderrama,
  Phys.\ Rev.\ D {\bf 88}, 054007 (2013)
  doi:10.1103/PhysRevD.88.054007
  [arXiv:1303.6608 [hep-ph]].

\bibitem{Wang:2013kva}
  P.~Wang and X.~G.~Wang,
  Phys.\ Rev.\ Lett.\  {\bf 111}, no. 4, 042002 (2013)
  doi:10.1103/PhysRevLett.111.042002
  [arXiv:1304.0846 [hep-ph]].

\bibitem{Wong:2003xk}
  C.~Y.~Wong,
  Phys.\ Rev.\ C {\bf 69}, 055202 (2004)
  doi:10.1103/PhysRevC.69.055202
  [hep-ph/0311088].

\bibitem{Swanson:2003tb}
  E.~S.~Swanson,
  Phys.\ Lett.\ B {\bf 588}, 189 (2004)
  doi:10.1016/j.physletb.2004.03.033
  [hep-ph/0311229].


\bibitem{Vijande:2004vt}
  J.~Vijande, F.~Fernandez and A.~Valcarce,
  Int.\ J.\ Mod.\ Phys.\ A {\bf 20}, 702 (2005)
  doi:10.1142/S0217751X05022214
  [hep-ph/0407136].

\bibitem{Maiani:2005pe}
  L.~Maiani, V.~Riquer, F.~Piccinini and A.~D.~Polosa,
  Phys.\ Rev.\ D {\bf 72}, 031502 (2005)
  doi:10.1103/PhysRevD.72.031502
  [hep-ph/0507062].

\bibitem{Navarra:2006nd}
  F.~S.~Navarra and M.~Nielsen,
  Phys.\ Lett.\ B {\bf 639}, 272 (2006)
  doi:10.1016/j.physletb.2006.06.054
  [hep-ph/0605038].

\bibitem{Cui:2006mp}
  Y.~Cui, X.~L.~Chen, W.~Z.~Deng and S.~L.~Zhu,
  HEPNP {\bf 31}, 7 (2007)
  [hep-ph/0607226].

\bibitem{Matheus:2006xi}
  R.~D.~Matheus, S.~Narison, M.~Nielsen and J.~M.~Richard,
  Phys.\ Rev.\ D {\bf 75}, 014005 (2007)
  doi:10.1103/PhysRevD.75.014005
  [hep-ph/0608297].

\bibitem{Nielsen:2006jn}
  M.~Nielsen, F.~S.~Navarra and M.~E.~Bracco,
  Braz.\ J.\ Phys.\  {\bf 37}, 56 (2007)
  doi:10.1590/S0103-97332007000100018
  [hep-ph/0609184].

\bibitem{Dubnicka:2010kz}
  S.~Dubnicka, A.~Z.~Dubnickova, M.~A.~Ivanov and J.~G.~Korner,
  Phys.\ Rev.\ D {\bf 81}, 114007 (2010)
  doi:10.1103/PhysRevD.81.114007
  [arXiv:1004.1291 [hep-ph]].

\bibitem{Dubnicka:2011mm}
  S.~Dubnicka, A.~Z.~Dubnickova, M.~A.~Ivanov, J.~G.~Koerner, P.~Santorelli and G.~G.~Saidullaeva,
  Phys.\ Rev.\ D {\bf 84}, 014006 (2011)
  doi:10.1103/PhysRevD.84.014006
  [arXiv:1104.3974 [hep-ph]].

\bibitem{Ebert:2005nc}
  D.~Ebert, R.~N.~Faustov and V.~O.~Galkin,
  Phys.\ Lett.\ B {\bf 634}, 214 (2006)
  doi:10.1016/j.physletb.2006.01.026
  [hep-ph/0512230].

\bibitem{Maiani:2004vq}
  L.~Maiani, F.~Piccinini, A.~D.~Polosa and V.~Riquer,
  Phys.\ Rev.\ D {\bf 71}, 014028 (2005)
  doi:10.1103/PhysRevD.71.014028
  [hep-ph/0412098].

\bibitem{Wang:2013vex}
  Z.~G.~Wang and T.~Huang,
  Phys.\ Rev.\ D {\bf 89}, no. 5, 054019 (2014)
  doi:10.1103/PhysRevD.89.054019
  [arXiv:1310.2422 [hep-ph]].

\bibitem{Close:2003mb}
  F.~E.~Close and S.~Godfrey,
  Phys.\ Lett.\ B {\bf 574}, 210 (2003)
  doi:10.1016/j.physletb.2003.09.011
  [hep-ph/0305285].

\bibitem{Li:2004sta}
  B.~A.~Li,
  Phys.\ Lett.\ B {\bf 605}, 306 (2005)
  doi:10.1016/j.physletb.2004.11.062
  [hep-ph/0410264].

\bibitem{Petrov:2005tp}
  A.~A.~Petrov,
  J.\ Phys.\ Conf.\ Ser.\  {\bf 9}, 83 (2005).
  doi:10.1088/1742-6596/9/1/013






\bibitem{LHCb:2020bwg}
R.~Aaij \textit{et al.} [LHCb],
Sci. Bull. \textbf{65} (2020) no.23, 1983-1993
doi:10.1016/j.scib.2020.08.032
[arXiv:2006.16957 [hep-ex]].



\bibitem{CMS}
K. Y. on behalf of the CMS Collaboration,  https:
//agenda.infn.it/event/28874/contributions/
170300/.


\bibitem{ATLAS}
E. B.-T. on behalf of the ATLAS Collabora-
tion, https://agenda.infn.it/event/28874/
contributions/170298/.







\bibitem{Albuquerque:2020hio}
R.~M.~Albuquerque, S.~Narison, A.~Rabemananjara, D.~Rabetiarivony and G.~Randriamanatrika,
Phys. Rev. D \textbf{102} (2020) no.9, 094001
doi:10.1103/PhysRevD.102.094001
[arXiv:2008.01569 [hep-ph]].

\bibitem{liu:2020eha}
M.~S.~liu, F.~X.~Liu, X.~H.~Zhong and Q.~Zhao,
[arXiv:2006.11952 [hep-ph]].


\bibitem{Lu:2020cns}
Q.~F.~L\"u, D.~Y.~Chen and Y.~B.~Dong,
Eur. Phys. J. C \textbf{80} (2020) no.9, 871
doi:10.1140/epjc/s10052-020-08454-1
[arXiv:2006.14445 [hep-ph]].




\bibitem{Giron:2020wpx}
J.~F.~Giron and R.~F.~Lebed,
Phys. Rev. D \textbf{102} (2020) no.7, 074003
doi:10.1103/PhysRevD.102.074003
[arXiv:2008.01631 [hep-ph]].


\bibitem{Dosch:2020hqm}
H.~G.~Dosch, S.~J.~Brodsky, G.~F.~de T\'eramond, M.~Nielsen and L.~Zou,
Nucl. Part. Phys. Proc. \textbf{312-317} (2021), 135-139
doi:10.1016/j.nuclphysbps.2021.05.035
[arXiv:2012.02496 [hep-ph]].


\bibitem{Yang:2020wkh}
B.~C.~Yang, L.~Tang and C.~F.~Qiao,
Eur. Phys. J. C \textbf{81} (2021) no.4, 324
doi:10.1140/epjc/s10052-021-09096-7
[arXiv:2012.04463 [hep-ph]].



\bibitem{Huang:2020dci}
G.~Huang, J.~Zhao and P.~Zhuang,
Phys. Rev. D \textbf{103} (2021) no.5, 054014
doi:10.1103/PhysRevD.103.054014
[arXiv:2012.14845 [hep-ph]].



\bibitem{Guo:2020pvt}
Z.~H.~Guo and J.~A.~Oller,
Phys. Rev. D \textbf{103} (2021) no.3, 034024
doi:10.1103/PhysRevD.103.034024
[arXiv:2011.00978 [hep-ph]].

\bibitem{Dong:2020nwy}
X.~K.~Dong, V.~Baru, F.~K.~Guo, C.~Hanhart and A.~Nefediev,
Phys. Rev. Lett. \textbf{126} (2021) no.13, 132001
[erratum: Phys. Rev. Lett. \textbf{127} (2021) no.11, 119901]
doi:10.1103/PhysRevLett.127.119901
[arXiv:2009.07795 [hep-ph]].


\bibitem{Gong:2020bmg}
C.~Gong, M.~C.~Du, Q.~Zhao, X.~H.~Zhong and B.~Zhou,
Phys. Lett. B \textbf{824} (2022), 136794
doi:10.1016/j.physletb.2021.136794
[arXiv:2011.11374 [hep-ph]].

\bibitem{Wang:2020wrp}
J.~Z.~Wang, D.~Y.~Chen, X.~Liu and T.~Matsuki,
Phys. Rev. D \textbf{103} (2021) no.7, 071503
doi:10.1103/PhysRevD.103.L071503
[arXiv:2008.07430 [hep-ph]].

\bibitem{Wang:2022jmb}
J.~Z.~Wang and X.~Liu,
[arXiv:2207.04893 [hep-ph]].


\bibitem{Gong:2022hgd}
C.~Gong, M.~C.~Du and Q.~Zhao,
[arXiv:2206.13867 [hep-ph]].


\bibitem{Liang:2022rew}
Z.~R.~Liang and D.~L.~Yao,
Rev. Mex. Fis. Suppl. \textbf{3} (2022) no.3, 0308042
doi:10.31349/SuplRevMexFis.3.0308042

\bibitem{Wan:2020fsk}
B.~D.~Wan and C.~F.~Qiao,
Phys. Lett. B \textbf{817} (2021), 136339
doi:10.1016/j.physletb.2021.136339
[arXiv:2012.00454 [hep-ph]].

\bibitem{Zhu:2020snb}
J.~W.~Zhu, X.~D.~Guo, R.~Y.~Zhang, W.~G.~Ma and X.~Q.~Li,
[arXiv:2011.07799 [hep-ph]].



\bibitem{Silvestre-Brac:1993zem}
B.~Silvestre-Brac and C.~Semay,
Z. Phys. C \textbf{57} (1993), 273-282
doi:10.1007/BF01565058


\bibitem{Chen:2016ont}
K.~Chen, X.~Liu, J.~Wu, Y.~R.~Liu and S.~L.~Zhu,
Eur. Phys. J. A \textbf{53} (2017) no.1, 5
doi:10.1140/epja/i2017-12199-3
[arXiv:1609.06117 [hep-ph]].

\bibitem{Weng:2021ngd}
X.~Z.~Weng, W.~Z.~Deng and S.~L.~Zhu,
Phys. Rev. D \textbf{105}, no.3, 034026 (2022)
doi:10.1103/PhysRevD.105.034026
[arXiv:2109.05243 [hep-ph]].


\bibitem{Jiang:2017tdc}
J.~F.~Jiang, W.~Chen and S.~L.~Zhu,
Phys. Rev. D \textbf{96} (2017) no.9, 094022
doi:10.1103/PhysRevD.96.094022
[arXiv:1708.00142 [hep-ph]].


\bibitem{Lu:2021kut}
Q.~F.~L\"u, D.~Y.~Chen, Y.~B.~Dong and E.~Santopinto,
Phys. Rev. D \textbf{104} (2021) no.5, 054026
doi:10.1103/PhysRevD.104.054026
[arXiv:2107.13930 [hep-ph]].



\bibitem{Valcarce:2005em}
A.~Valcarce, H.~Garcilazo, F.~Fernandez and P.~Gonzalez,
Rept. Prog. Phys. \textbf{68} (2005), 965-1042
doi:10.1088/0034-4885/68/5/R01
[arXiv:hep-ph/0502173 [hep-ph]].

\bibitem{Segovia:2008zz}
J.~Segovia, A.~M.~Yasser, D.~R.~Entem and F.~Fernandez,
Phys. Rev. D \textbf{78} (2008), 114033
doi:10.1103/PhysRevD.78.114033

\bibitem{Segovia:2010zzb}
J.~Segovia, D.~R.~Entem and F.~Fernandez,
J. Phys. G \textbf{37} (2010), 075010
doi:10.1088/0954-3899/37/7/075010

\bibitem{Segovia:2011tb}
J.~Segovia, D.~R.~Entem, F.~Fernandez and E.~Ruiz Arriola,
Phys. Rev. D \textbf{85} (2012), 074001
doi:10.1103/PhysRevD.85.074001
[arXiv:1108.0208 [hep-ph]].

\bibitem{Segovia:2016xqb}
J.~Segovia, P.~G.~Ortega, D.~R.~Entem and F.~Fern\'andez,
Phys. Rev. D \textbf{93} (2016) no.7, 074027
doi:10.1103/PhysRevD.93.074027
[arXiv:1601.05093 [hep-ph]].
\bibitem{Segovia:2011dg}
J.~Segovia, C.~Albertus, D.~R.~Entem, F.~Fernandez, E.~Hernandez and M.~A.~Perez-Garcia,
Phys. Rev. D \textbf{84} (2011), 094029
doi:10.1103/PhysRevD.84.094029
[arXiv:1107.4248 [hep-ph]].

\bibitem{Segovia:2011zza}
J.~Segovia, D.~R.~Entem and F.~Fernandez,
Phys. Rev. D \textbf{83} (2011), 114018
doi:10.1103/PhysRevD.83.114018
\bibitem{Segovia:2012cd}
J.~Segovia, D.~R.~Entem and F.~Fern\'andez,
Phys. Lett. B \textbf{715} (2012), 322-327
doi:10.1016/j.physletb.2012.08.005
[arXiv:1205.2215 [hep-ph]].
\bibitem{Segovia:2013kg}
J.~Segovia, D.~R.~Entem and F.~Fernandez,
Nucl. Phys. A \textbf{915} (2013), 125-141
doi:10.1016/j.nuclphysa.2013.07.004
[arXiv:1301.2592 [hep-ph]].
\bibitem{Segovia:2014mca}
J.~Segovia, D.~R.~Entem and F.~Fern\'andez,
Phys. Rev. D \textbf{91} (2015) no.1, 014002
doi:10.1103/PhysRevD.91.014002
[arXiv:1409.7079 [hep-ph]].
\bibitem{Ortega:2009hj}
P.~G.~Ortega, J.~Segovia, D.~R.~Entem and F.~Fernandez,
Phys. Rev. D \textbf{81} (2010), 054023
doi:10.1103/PhysRevD.81.054023
[arXiv:0907.3997 [hep-ph]].


\bibitem{Ortega:2016hde}
P.~G.~Ortega, J.~Segovia, D.~R.~Entem and F.~Fern\'andez,
Phys. Rev. D \textbf{94} (2016) no.11, 114018
doi:10.1103/PhysRevD.94.114018
[arXiv:1608.01325 [hep-ph]].

\bibitem{Jin:2020yjn}
X.~Jin, Y.~Xue, H.~Huang and J.~Ping,
Eur. Phys. J. C \textbf{80} (2020) no.11, 1083
doi:10.1140/epjc/s10052-020-08650-z
[arXiv:2006.13745 [hep-ph]].

\bibitem{Yan:2021glh}
Y.~Yan, Y.~Wu, X.~Hu, H.~Huang and J.~Ping,
Phys. Rev. D \textbf{105} (2022) no.1, 014027
doi:10.1103/PhysRevD.105.014027
[arXiv:2110.10853 [hep-ph]].

\bibitem{Liu:2020yen}
X.~Liu, H.~Huang, J.~Ping and D.~Chen,
Phys. Rev. C \textbf{103} (2021) no.2, 025202
doi:10.1103/PhysRevC.103.025202
[arXiv:2010.15398 [hep-ph]].

\bibitem{Jin:2020jfc}
X.~Jin, Y.~Xue, H.~Huang and J.~Ping,
Eur. Phys. J. C \textbf{80} (2020) no.11, 1083
doi:10.1140/epjc/s10052-020-08650-z
[arXiv:2006.13745 [hep-ph]].


\bibitem{Liu:2018nse}
X.~Liu, H.~Huang and J.~Ping,
Phys. Rev. C \textbf{98} (2018) no.5, 055203
doi:10.1103/PhysRevC.98.055203
[arXiv:1807.03195 [hep-ph]].



\bibitem{Liu:2022vyy}
X.~Liu, D.~Chen, H.~Huang, J.~Ping, X.~Chen and Y.~Yang,
[arXiv:2204.08104 [hep-ph]].

\bibitem{Kamimura:1981oxj}
M.~Kamimura,
Nucl. Phys. A \textbf{351} (1981), 456-480
doi:10.1016/0375-9474(81)90182-2

\bibitem{ParticleDataGroup:2018ovx}
M.~Tanabashi \textit{et al.} [Particle Data Group],
Phys. Rev. D \textbf{98} (2018) no.3, 030001
doi:10.1103/PhysRevD.98.030001

\bibitem{Huang:2013rla}
H.~Huang, J.~Ping and F.~Wang,
Phys. Rev. C \textbf{89} (2014) no.3, 035201
doi:10.1103/PhysRevC.89.035201
[arXiv:1311.4732 [hep-ph]].

\end{thebibliography}
\end{document}